\documentclass[aps,pre,amsmath,amssymb,longbibliography,lengthcheck,showpacs,superscriptaddress,floatfix]{revtex4-1}
\usepackage{graphicx,color}
\def\vec#1{\mbox{\boldmath $#1$}}

\begin{document}
\title{Elastic Moduli and Vibrational Modes in Jammed Particulate Packings}

\author{Hideyuki Mizuno}
\email{E-mail: hideyuki.mizuno@fukui.kyoto-u.ac.jp} 
\altaffiliation{Current address: Fukui Institute for Fundamental Chemistry, Kyoto University, Kyoto 606-8103, Japan}
\affiliation{Institut f\"{u}r Materialphysik im Weltraum, Deutsches Zentrum f\"{u}r Luft- und Raumfahrt (DLR), 51170 K\"{o}ln, Germany}

\author{Kuniyasu Saitoh}
\altaffiliation{Current address: WPI-Advanced Institute for Materials Research, Tohoku University, Sendai 980-8577, Japan}
\affiliation{Faculty of Engineering Technology, MESA+, University of Twente, 7500 AE Enschede, The Netherlands}

\author{Leonardo E.~Silbert}
\affiliation{Department of Physics, Southern Illinois University Carbondale, Carbondale, Illinois 62901, USA}

\date{\today}

\begin{abstract}
When we elastically impose a homogeneous, affine deformation on amorphous solids, they also undergo an inhomogeneous, non-affine deformation, which can have a crucial impact on the overall elastic response.
To correctly understand the elastic modulus $M$, it is therefore necessary to take into account not only the affine modulus $M_A$, but also the non-affine modulus $M_N$ that arises from the non-affine deformation.
In the present work, we study the bulk ($M=K$) and shear ($M=G$) moduli in static jammed particulate packings over a range of packing fractions $\varphi$.
The affine $M_A$ is determined essentially by the static structural arrangement of particles, whereas the non-affine $M_N$ is related to the vibrational eigenmodes.
One novelty of this work is to elucidate the contribution of each vibrational mode to the non-affine $M_N$ through a modal decomposition of the displacement and force fields.
In the vicinity of the (un)jamming transition, $\varphi_{c}$, the vibrational density of states, $g(\omega)$, shows a plateau in the intermediate frequency regime above a characteristic frequency $\omega^\ast$.
We illustrate that this unusual feature apparent in $g(\omega)$ is reflected in the behavior of $M_N$: As $\varphi \rightarrow \varphi_c$, where $\omega^\ast \rightarrow 0$, those modes for $\omega < \omega^\ast$ contribute less and less, while contributions from those for $\omega > \omega^\ast$ approach a constant value which results in $M_N$ to approach a critical value $M_{Nc}$, as $M_N-M_{Nc} \sim \omega^\ast$.
At $\varphi_c$ itself, the bulk modulus attains a finite value $K_c=K_{Ac}-K_{Nc} > 0$, such that $K_{Nc}$ has a value that remains below $K_{Ac}$.
In contrast, for the critical shear modulus $G_c$, $G_{Nc}$ and $G_{Ac}$ approach the same value so that the total value becomes exactly zero, $G_c = G_{Ac}-G_{Nc} =0$.
We explore what features of the configurational and vibrational properties cause such the distinction between $K$ and $G$, allowing us to validate analytical expressions for their critical values.
\end{abstract}

\maketitle

\section{Introduction}
A theoretical foundation to determine and predict the elastic response of amorphous solids persists as an ongoing problem in the soft condensed matter community~\cite{Alexander_1998}.
As developed, the classical theory of linear elasticity of solids is based on the concept of affineness~\cite{elastictheory2,elastictheory,Ashcroft,kettel}: The elastic response of solids is inferred on assuming an affine deformation, i.e., the constituent particles are assumed to follow the imposed, homogeneous, affine deformation field.
For that case, the elastic modulus can be formulated through the so-called Born-Huang expression, which we denote as the affine modulus in this paper.
In contrast, amorphous solids, such as molecular, polymer, and colloidal glasses~\cite{monaco_2009,monaco2_2009,tsamados_2009,Fan_2014,Wagner_2011,Hufnagel_2015,Mayr_2009,Mizuno_2013,Wittmer_2013,Wittmer2_2013,Wittmer_2015,yoshimoto_2004,Zaccone_2013,makke_2011,Klix_2012}, disordered crystals~\cite{Kaya_2010,Mizuno2_2013,Mizuno_2014}, and athermal jammed or granular packings~\cite{OHern_2002,OHern_2003,Ellenbroek_2006,Ellenbroek_2009,Ellenbroek2_2009,Wyart,Maloney_2004,Maloney2_2006,Maloney_2006,Lemaitre_2006,Karmakar_2010,Hentschel_2011,Zaccone_2011,Zaccone2_2011,Zaccone_2014,Lerner_2014,Karimi_2015}, exhibit inhomogeneous, non-affine deformations or relaxations, which cause the system to deviate from the homogeneous affine state, significantly impacting the elastic response.
In such cases, the Born-Huang expression for the elastic modulus requires the addition of non-negligible correction arising from the non-affine deformation.
Therefore, the key to determining the mechanical properties of amorphous solids lies in understanding the role played by their non-affine response~\cite{Makse_1999,Wittmer_2002,Tanguy_2002,leonforte_2005,DiDonna_2005}.
Here, it should be noted that the presence of disorder is not the only defining property necessary for observing non-affine behavior.
While a perfectly ordered crystalline solid with a single atom per unit cell shows a true affine response, such that the Born-Huang expression becomes exact in this case, crystals with a multi-atom unit cell generally exhibit non-affine responses~\cite{Jaric_1988}.
Thus, investigating the fundamental mechanisms that lead to non-affine behavior is a topic of interest to the broader community concerned with materials characterization.

When all the constituent particles in an amorphous solid are displaced according to a homogeneous affine strain field, its immediate elastic response is described by the affine deformation with its associated, affine modulus (or the Born-Huang expression)~\cite{elastictheory,elastictheory2,Ashcroft,kettel,Alexander_1998}.
However, due to the amorphous structure, whereby the local environment of each particle is slightly different from every other particle, the imposed affine deformation actually causes the forces on individual particles to become unbalanced in a heterogeneous manner~\cite{Maloney_2004,Maloney2_2006,Maloney_2006,Lemaitre_2006}.
Thus, as the particles seek pathways to relax back towards a new state of mechanical equilibrium, they adopt a configuration that is different from the originally imposed affine deformation field~\cite{Wittmer_2002,Tanguy_2002,leonforte_2005,DiDonna_2005,Maloney_2004,Maloney2_2006,Maloney_2006,Lemaitre_2006}.
Consequently, the elastic response of an amorphous solid cannot be described by the affine deformation response alone.
It also becomes necessary to take into account the non-affine deformation (relaxation).
The elastic modulus is therefore composed of two components~\cite{Mayr_2009,Mizuno_2013,Wittmer_2013,Wittmer2_2013,Wittmer_2015,yoshimoto_2004,Zaccone_2013,Mizuno2_2013,Mizuno_2014,Maloney_2004,Maloney2_2006,Maloney_2006,Lemaitre_2006,Karmakar_2010,Hentschel_2011,Zaccone_2011,Zaccone2_2011,Zaccone_2014}:
(i) The affine modulus, which comes from the imposed affine deformation, and (ii) the non-affine modulus, which is considered as an energy dissipation term during non-affine relaxation, or more specifically regarded as a inhomogeneous repartitioning of the interaction potential energy during the relaxation process as work done along the non-affine pathways.

In the harmonic limit, the affine modulus essentially derives directly from the static configuration of the constituent particles and the interaction potential between them.
Whereas, the non-affine modulus is formulated in terms of the vibrational eigenmodes (eigenvalues and eigenvectors) of the system~\cite{Lutsko_1989,Maloney_2004,Maloney2_2006,Maloney_2006,Lemaitre_2006,Karmakar_2010,Hentschel_2011,Zaccone_2011,Zaccone2_2011,Zaccone_2014}, which can be obtained by performing a normal mode analysis on the dynamical matrix~\cite{Ashcroft,kettel,McGaughey}.
Physically this means that the vibrational eigenmodes are excited during the non-affine deformation process, contributing to the energy relaxation (the non-affine elastic modulus)~\cite{Wittmer_2002,Tanguy_2002}.
In this sense, the nonaffine modulus can be constructed as a product of the inherent displacement field and corresponding force field~\cite{Maloney_2004,Maloney2_2006,Maloney_2006,Lemaitre_2006}, which are defined through the eigenmodes.
Thus, we expect that any unusual features expressed by the vibrational properties of amorphous solids should be reflected in their elastic properties.
Indeed, it is well known that (both thermal and athermal) amorphous materials exhibit anomalous features in their vibrational states, such as an excess of low-frequency modes (Boson peak)~\cite{monaco_2009,monaco2_2009,Kaya_2010,Mizuno2_2013,Mizuno_2014} and localizations of modes~\cite{mazzacurati_1996,Allen_1999,Schober_2004,Silbert_2009,Xu_2010}, which should be reflected in the behavior of the non-affine modulus.
In addition, Maloney and Lema\^{i}tre~\cite{Maloney_2004,Maloney2_2006} demonstrated that at the onset of a plastic event in an overcompressed disc packing under shear, a single eigenmode frequency goes to zero, which causes the non-affine modulus to diverge (toward $-\infty$) initiating the plastic event.

A paradigmatic system that expresses the generic features of amorphous materials is the case of an isotropically, overcompressed, static, jammed packing of particles~\cite{OHern_2002,OHern_2003,Ellenbroek_2006,Ellenbroek_2009,Ellenbroek2_2009}.
As we decompress the jammed system, it unjams - goes from solid to fluid phase - at a particular packing fraction of particles, $\varphi_c$, that is the unjamming transition.
The jamming (unjamming) point, $\varphi_{c}$, signals the transition between a mechanically robust solid phase and a collection of non-contacting particles unable to support mechanical perturbations.
In such athermal solids, peculiar vibrational features are readily apparent in the vibrational density of states (vDOS), $g(\omega)$~\cite{Silbert_2009,Xu_2010,Silbert_2005}.
The vDOS exhibits a plateau in the intermediate frequency regime, $\omega > \omega^\ast$, above some characteristic frequency $\omega^\ast$~(see also Fig.~\ref{fig.vibration}(a)).
On approach to the transition point $\varphi_c$, this plateau regime extends down to zero frequency, as the onset frequency $\omega^\ast$ goes to zero, $\omega^\ast \rightarrow 0$~\cite{Silbert_2005}.
Wyart \textit{et. al.}~\cite{Wyart_2005,Wyart_2006,Xu_2007} described the vibrational modes in the plateau regime of $g(\omega)$, in terms of ``anomalous'' modes emerging from the isostatic feature of marginally stable packings.
More recent work~\cite{Goodrich_2013} proposed an alternative description based on the concept of a rigidity length scale.
Either way, the progressive development of vibrational modes in the plateau regime seems to play a crucial role in controlling the mechanical properties of marginally jammed solids, e.g., in the loss of rigidity at the transition $\varphi_c$.

\begin{figure*}[t]
\centering
\includegraphics[width=0.98\textwidth]{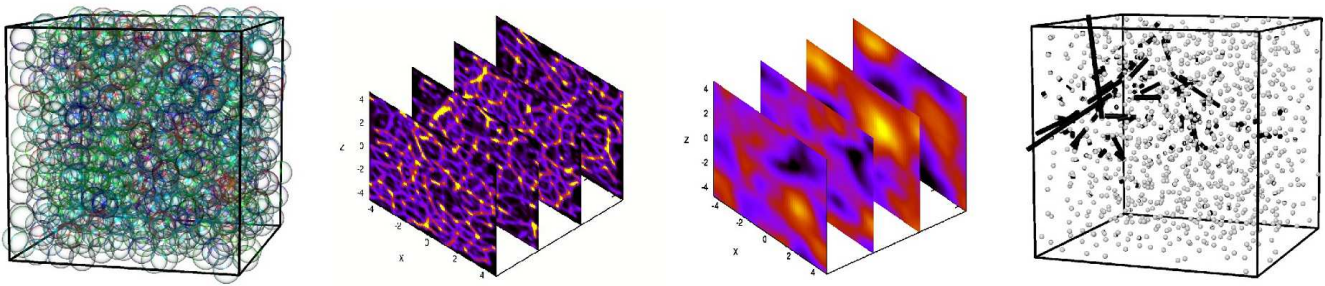}
\vspace*{0mm}
\caption{\label{fig.panel}
(Color online) An aper\c cu of the study presented here.
From left to right: Particle configuration, spatial maps of the coarse-grained normal forces within the packing, maps of the local shear modulus, and a sample eigenvector.
This representative configuration consists of a static packing of monodisperse spheres prepared at a packing fraction, $\Delta \varphi = 10^{-6}$ (shading represents particle coordination number).
The force network and shear modulus maps are collages of slices approximately one and three particle diameter thick through a packing, respectively.
The eigenvector panel emphasizes the individual polarization vector of each particle in a typical high-frequency, localized mode ($\omega = 2.5$) (particle centers represented by small spheres).}
\end{figure*}

In the present work, by using a model jammed packing of particles interacting via a finite-range, repulsive potential (see Fig.~\ref{fig.panel} and Eq.~(\ref{interaction})), we study the compressive, bulk modulus $K$ and the shear modulus $G$, close to the transition point $\varphi_c$.
We execute a comprehensive analysis of the affine and non-affine components of these two elastic moduli.
A main novelty of the present work is to elucidate the contribution to the non-affine moduli, from each vibrational mode, particularly those in the plateau regime of $g(\omega)$.
To achieve this, we perform a normal mode analysis of the dynamical matrix~\cite{Ashcroft,kettel,McGaughey}, and then an eigenmode decomposition of the non-affine moduli~\cite{Lutsko_1989,Maloney_2004,Maloney2_2006,Maloney_2006,Lemaitre_2006,Karmakar_2010,Hentschel_2011,Zaccone_2011,Zaccone2_2011,Zaccone_2014}.
Thereby, we avoid the need to explicitly apply a deformation to the packings which can be troublesome for very fragile systems close to $\varphi_{c}$.
We demonstrate that in the plateau regime above $\omega^\ast$, each vibrational mode similarly contributes to the non-affine elastic moduli, i.e., the contribution is independent of the eigenmode frequency.
This behavior derives from the competing influences of the displacement and force fields that are in turn largely set by low-frequency modes and high-frequency modes, respectively.
In addition, the modal contribution shows a crossover at $\omega^\ast$, from the plateau independence for $\omega > \omega^\ast$, to a growing behavior $\sim \omega^{-2}$ (with decreasing $\omega$) for $\omega < \omega^\ast$.
We show that this crossover at $\omega^\ast$ is controlled by the competition between compressing/stretching and sliding vibrational energies.

As the system approaches the unjamming transition from above, and passes into the fluid phase, the two elastic moduli, $K$ and $G$, show distinct critical behaviors: The bulk modulus $K$ discontinuously drops to zero, whereas the shear modulus $G$ continuously goes to zero, $G \rightarrow 0$~\cite{OHern_2002,OHern_2003,Ellenbroek_2006,Ellenbroek_2009,Ellenbroek2_2009,Wyart}.
At the transition itself of the packing fraction $\varphi_c$, the critical value of the affine component of the bulk modulus remains above that of the nonaffine counterpart, whence the total modulus $K$ takes on a finite, positive value.
In contrast, for the shear modulus, the non-affine modulus cancels out the affine modulus, leading to the shear modulus becoming identically zero at the transition.
Here, we explore what features in the configurational and vibrational properties of jammed solids cause such the distinction between these critical behaviors, which leads us to derive the critical values of $K$ and $G$, analytically.
An overview of our study is shown in Fig.~\ref{fig.panel}.

The rest of this paper is organized as follows.  In Sec.~\ref{NM} we outline the simulation method.
We describe the system of jammed packings and the method for vibrational eigenmode analysis.
We also discuss in detail the linear response formulation for obtaining the linear elastic moduli and their modal decomposition.
Section \ref{results} contains a comprehensive presentation of our results.
This section is broken down into several subsections that focus on the affine and non-affine moduli, characterization of the eigenmodes themselves, the modal contributions to elastic moduli, and derivations of the critical values of the elastic moduli.
We summarize our results in Sec~\ref{summary}, and end with an extensive set of conclusive remarks in Sec.~\ref{conclusions}.

\section{Numerical method} \label{NM}
\subsection{System description}
We study a 3-dimensional ($d=3$) athermal jammed solid, which is composed of mono-disperse, frictionless, deformable particles with diameter $\sigma$ and mass $m$.
Configurations of static, mechanically stable states are prepared over a wide range in packing pressure in a cubic simulation box with periodic boundary conditions in all three ($x,y,z$) directions, using a compression/decompression protocol~\cite{Silbert_2010} implemented within the open-source, molecular dynamics package LAMMPS~\cite{plimpton_1995}.
Particles, $i$ and $j$, interact via a finite-range, purely repulsive, harmonic potential;
\begin{equation} \label{interaction}
\phi(r_{ij}) =
\left\{ \begin{aligned}
& \frac{ \textrm{k} }{2} \left(\sigma-r_{ij} \right)^2 & (r_{ij} < \sigma), \\
& 0 & (r_{ij} \ge \sigma),
\end{aligned} \right. \\
\end{equation}
where $r_{ij}=|\vec{r}_i-\vec{r}_j|$ is the distance between particles $i$ and $j$, the $\vec{r}_i$ is particle position vector, and $\textrm{k}$ parameterizes the particle stiffness and sets an energy scale through $\textrm{k}\sigma^{2}$.
In the following, we use $\sigma$, $m$, and $\tau=(m/\textrm{k})^{1/2}$ as units of length, mass, and time, respectively, i.e., we set $\sigma=m=\textrm{k}=1$.

When $r_{ij} < \sigma$, the pair of particles, $(i,j)$, feels a finite potential, i.e., particles are connected.
In the present study, we always removed rattler particles which have less than $3$ contacting neighbors, and the total number of particles is $N \simeq 1000$ (precise number $N$ depends on the configuration realizations that we used to average our data).
We denote the number of connected pairs of particles as $N^\text{ct}=Nz/2$, where $z$ is the average contact number per particle (or the coordination number).
At the transition point $\varphi_c$, where the system is in the isostatic state~\cite{Wyart,Wyart_2005,Wyart_2006,Maxwell_1864}, the number of connections (constraints) is precisely balanced by the number of degrees of freedom, i.e., $N^\text{ct}_c=3N-3$ (three ($x,y,z$) translational degrees of freedom are removed), and the contact number is
\begin{equation}
z_c = \frac{2N^\text{ct}_c}{N} = 6 \left( 1- \frac{1}{N} \right),
\end{equation}
which is $6=2d$ in the thermodynamic limit, $N \rightarrow \infty$.
The total potential energy $E$ of the system is then given by (using $\sigma = \textrm{k} = 1$)
\begin{equation} \label{energy1}
E = \sum_{(i,j)} \phi(r_{ij}) = \sum_{(i,j)} \frac{1}{2} \left(1-{r_{ij}} \right)^2,
\end{equation}
where the summation, $\sum_{(i,j)}$, runs over all connected pairs of particles, $(i,j) \in N^\text{ct}$.

The temperature is zero, $T=0$, and the packing fraction of particles, $\varphi$, is the control parameter that we use to systematically probe static packings of varying rigidity~\cite{OHern_2002,OHern_2003,Ellenbroek_2006,Ellenbroek_2009,Ellenbroek2_2009};
\begin{equation}
\varphi = \frac{\pi N}{6V} = \frac{\pi}{6} \hat{\rho},
\end{equation}
where $V=L^3$ is the total volume ($L$ is the system length), and $\hat{\rho} = N/V$ is the number density.
The critical value of $\varphi$ at the transition is found to coincide with the value of random close packing, $\varphi_c \simeq 0.64$, in $d=3$ dimensions~\cite{OHern_2002,OHern_2003}.
The critical value of $\hat{\rho}$ is then given as $\hat{\rho}_c = (6/\pi)\varphi_c \simeq 1.2$.
We study the jammed solid phase above the transition point $\varphi_c$, and characterize the rigidity of the system by the distance from $\varphi_c$, i.e., $\Delta \varphi = \varphi - \varphi_c \ge 0$.
In the present work, we varied $\Delta \varphi$ by five decades, $10^{-6} \le \Delta \varphi \le 10^{-1}$.
At each $\Delta \varphi$, $100$ configuration realizations were prepared, and the values of quantities were obtained by averaging over those $100$ realizations.

\subsection{Unstressed system}
In the harmonic limit, the energy variation, $\delta E$, due to the displacements of particles from the equilibrium positions $\{\vec{r}_1,\vec{r}_2,...,\vec{r}_N \}$ by $\{\delta \vec{R}_{1},\delta \vec{R}_{2},...,\delta \vec{R}_{N} \}$ is formulated as~\cite{Ellenbroek_2006,Ellenbroek_2009,Ellenbroek2_2009,Wyart_2005,Wyart_2006,Xu_2007}
\begin{equation}
\begin{aligned} \label{energy2}
\delta E &= \sum_{(i,j)} \left[ \frac{\phi''(r_{ij})}{2} {\delta \vec{R}_{ij}^{\parallel}}^2 + \frac{\phi'(r_{ij})}{2 r_{ij}} {\delta \vec{R}_{ij}^{\perp}}^2 \right], \\
&:= \delta E^{\parallel} - \delta E^{\perp},
\end{aligned}
\end{equation}
where $\phi'(r_{ij})$ and $\phi''(r_{ij})$ are respectively the first and second derivatives of the potential $\phi(r_{ij})$ with respect to $r_{ij}$.
The vectors, $\delta \vec{R}_{ij}^{\parallel}$ and $\delta \vec{R}_{ij}^{\perp}$, are projections of $\delta \vec{R}_{ij} = \delta \vec{R}_{i}-\delta \vec{R}_{j}$ onto the planes parallel and perpendicular to $\vec{r}_{ij}=\vec{r}_{i}-\vec{r}_{j}$ (the equilibrium separation vector), respectively;
\begin{equation}
\begin{aligned} \label{energy3}
\delta \vec{R}_{ij}^{\parallel} &= \left(\delta \vec{R}_{ij} \cdot {\vec{n}_{ij}} \right) \vec{n}_{ij}, \\
\delta \vec{R}_{ij}^{\perp} &= \delta \vec{R}_{ij} - \left(\delta \vec{R}_{ij} \cdot {\vec{n}_{ij}} \right) \vec{n}_{ij},
\end{aligned}
\end{equation}
with $\vec{n}_{ij} = {\vec{r}_{ij}}/{{r}_{ij}}$, the unit vector of $\vec{r}_{ij}$.
In the present paper, we call $\vec{n}_{ij}$ the ``bond vector" of contact $(i,j)$.
As in Eq.~(\ref{energy2}), $\delta E$ is decomposed into two terms, $\delta E^{\parallel}\ (\ge 0)$ and $-\delta E^{\perp}\ (\le 0)$, which are energy variations due to compressing/stretching motions, $\delta \vec{R}_{ij}^{\parallel}$, and transverse sliding motions, $\delta \vec{R}_{ij}^{\perp}$, respectively~\cite{Ellenbroek_2006,Ellenbroek_2009,Ellenbroek2_2009,Wyart_2005,Wyart_2006,Xu_2007}.

In the jammed solid state $\Delta \varphi > 0$, the pressure $p > 0$ is finite (positive), and the first derivative of the potential, $\phi'(r_{ij})$, which corresponds to the contact force, is a finite (negative) value between the connected pair of particles, $(i,j)$.
For this reason we refer to such a state as the ``stressed'' state.
Besides this original stressed system, we have also studied the ``unstressed'' system~\cite{Wyart_2005,Wyart_2006,Xu_2007}, where we keep the second derivative $\phi''(r_{ij})$ but drop the first derivative $\phi'(r_{ij}) \equiv 0$, i.e., we replace stretched springs between connected particles by unstretched (relaxed) springs of the same stiffness $\phi''(r_{ij})$.
Note that the unstressed system is stable to keep exactly the same configuration of the original stressed system, with zero pressure, $p=0$.
In the stressed system, the sliding motion $\delta \vec{R}_{ij}^{\perp}$ reduces the potential energy by $\delta E^{\perp} >0$ (see Eq.~(\ref{energy2})) and destabilizes the system~\cite{Ellenbroek_2006,Ellenbroek_2009,Ellenbroek2_2009,Wyart_2005,Wyart_2006,Xu_2007}, whereas $\delta \vec{R}_{ij}^{\perp}$ in the unstressed system does \textit{not} contribute to the energy variation, i.e., $\delta E^{\perp} \equiv 0$.
Thus, by comparing the stressed and unstressed systems, we can separately investigate the effects of these two types of motions, the normal $\delta \vec{R}_{ij}^{\parallel}$ and tangential $\delta \vec{R}_{ij}^{\perp}$ motions, on energy-related quantities such as the elastic moduli.

\subsection{Vibrational eigenmodes} \label{sec.vibration}
The vibrational eigenmodes are obtained by means of the standard normal mode analysis~\cite{Ashcroft,kettel,McGaughey}.
We have solved the eigenvalue problem of the dynamical matrix $H$,
\begin{equation}
H = \frac{\partial^2 E}{\partial\vec{r} \partial \vec{r}} = \left[ \frac{\partial^2 E}{\partial\vec{r}_i \partial \vec{r}_j} \right]_{i,j=1,2,...,N},
\end{equation}
with $\vec{r}=\left[\vec{r}_1,\vec{r}_2,...,\vec{r}_N \right]$, in order to get the eigenvalues, $\lambda^k$, and the eigenvectors, $\vec{e}^k=\left[\vec{e}^{k}_1,\vec{e}^{k}_2,...,\vec{e}^{k}_N \right]$, for vibrational modes $k=1,2,...,3N-3$ (the three ($x,y,z$) zero-frequency translational modes are removed).
Note that $\vec{r}$ and $\vec{e}^k$ are $3N$ dimensional vectors, and $H$ is the $3N \times 3N$ Hessian matrix.
Since we always remove rattler particles, there are no zero-frequency modes associated with them, thus $3N-3$ eigenvalues are all positive-definite, $\lambda^k > 0$.

The quantity, $\omega^k = \sqrt{\lambda^k}$, is the eigenfrequency of the mode $k$~\cite{Ashcroft,kettel,McGaughey}, from which we calculate the vDOS $g(\omega)$;
\begin{equation} \label{vdoseq} 
g(\omega) = \frac{1}{3N-3} \sum_{k=1}^{3N-3} \delta \left( \omega-\omega^k \right),
\end{equation}
where $\delta(x)$ is the Dirac delta function.
The eigenvector $\vec{e}^k=\left[\vec{e}^{k}_1,\vec{e}^{k}_2,...,\vec{e}^{k}_N \right]$, which is normalized as $\vec{e}^k \cdot \vec{e}^l = \sum_{i=1}^N \vec{e}^k_i \cdot \vec{e}^l_i = \delta_{kl}$ ($\delta_{kl}$ is the Kronecker delta), is the polarization field of particles in mode $k$, i.e., each particle $i$ ($=1,2,...,N$) vibrates along its polarization vector $\vec{e}^k_i$.
The vector, $\vec{e}_{ij}^k=\vec{e}_{i}^k-\vec{e}_{j}^k$, represents the vibrational motion between particle pair, $(i,j)$.
Like $\delta \vec{R}_{ij}$ in Eq.~(\ref{energy3}), $\vec{e}_{ij}^k$ can also be decomposed into the normal $\vec{e}_{ij}^{k \parallel}$ and tangential $\vec{e}_{ij}^{k \perp}$ vibrational motions with respect to the connecting bond vector $\vec{n}_{ij}$;
\begin{equation}
\begin{aligned} \label{vs2}
\vec{e}_{ij}^{k \parallel} &= \left( \vec{e}^k_{ij} \cdot {\vec{n}_{ij}} \right) \vec{n}_{ij}, \\
\vec{e}_{ij}^{k \perp} &= \vec{e}^k_{ij} - \left( \vec{e}^k_{ij} \cdot {\vec{n}_{ij}} \right) \vec{n}_{ij}.
\end{aligned}
\end{equation}

By substituting $\vec{e}_{ij}^{k \parallel}$ and $\vec{e}_{ij}^{k \perp}$ into $\delta \vec{R}_{ij}^{\parallel}$ and $\delta \vec{R}_{ij}^{\perp}$ in Eq.~(\ref{energy2}), we obtain the vibrational energy $\delta E^k$ of the mode $k$;
\begin{equation}
\begin{aligned} \label{vs1}
\delta E^k &= \sum_{(i,j)} \left[ \frac{\phi''(r_{ij})}{2} {\vec{e}_{ij}^{k \parallel}}^2 + \frac{\phi'(r_{ij})}{2 r_{ij}} {\vec{e}_{ij}^{k \perp}}^2 \right], \\
&:= \delta E^{k \parallel} - \delta E^{k \perp}.
\end{aligned}
\end{equation}
$\delta E^{k \parallel}\ (\ge 0)$ and $-\delta E^{k \perp}\ (\le 0)$ are energies due to the compressional/stretching, $\vec{e}_{ij}^{k \parallel}$, and sliding, $\vec{e}_{ij}^{k \perp}$, vibrational motions, respectively.
$\delta E^k$ is also formulated as~\cite{Ashcroft,kettel,McGaughey}
\begin{equation} \label{vs3}
\delta E^k = \frac{1}{2} \left( \vec{e}^k \cdot H \cdot \vec{e}^k \right) = \frac{ \lambda^k }{2} = \frac{ {\omega^k}^2 }{2}.
\end{equation}
Thus, Eqs.~(\ref{vs1}) and~(\ref{vs3}) give
\begin{equation} \label{vs4}
\sum_{(i,j)} \left[ \phi''(r_{ij}) {\vec{e}_{ij}^{k \parallel}}^2 + \frac{\phi'(r_{ij})}{r_{ij}} {\vec{e}_{ij}^{k \perp}}^2 \right] = {\omega^k}^2.
\end{equation}

In the present work, we characterize the vibrational mode $k$ in terms of the quantities described above, i.e., $\omega^k, \vec{e}_{ij}^{k \parallel}, \vec{e}_{ij}^{k \perp}, \delta E^{k \parallel}, \delta E^{k \perp}$, which will be presented in Sec.~\ref{sec.vibrationalstate}.
We note that those quantities are different between the original stressed system and the unstressed system, since the dynamical matrix is different between them.
In Sec.~\ref{sec.vibrationalstate}, we will also compare the vibrational modes between the two systems.

\subsection{Elastic moduli} \label{sec.moduli}
The linear elastic response of the isotropic systems studied here is characterized by two elastic moduli: The bulk modulus $K$ is for volume-changing bulk deformation ${\epsilon}_K$, and the shear modulus $G$ for volume-preserving shear deformation ${\epsilon}_G$, where ${\epsilon}_K$ and ${\epsilon}_G$ are the strains representing the global affine deformations~\cite{elastictheory2,elastictheory,Ashcroft,kettel,Alexander_1998}.
In the present paper, we represent $M$ for those two elastic moduli, i.e., $M=K, G$.
Rather than explicitly applying a deformation field to the systems at hand, we calculate the elastic modulus $M$ through the harmonic formulation, which has been established and employed in previous studies~\cite{Lutsko_1989,Maloney_2004,Maloney2_2006,Maloney_2006,Karmakar_2010,Lemaitre_2006,Hentschel_2011,Zaccone_2011,Zaccone2_2011,Zaccone_2014}.
In the following, we introduce the formulation and notations for modulus $M$.
We show the formulation of only $C_{xyxy}$ (Voigt notation) for the shear modulus $G$, but the other shear moduli, e.g., $C_{xzxz}, C_{yzyz}$, coincide with $C_{xyxy}$ in the isotropic system and give the same results.

As we described in the introduction, the elastic modulus, $M=K,G$, has two components, the affine modulus, $M_A = K_A,G_A$, and the non-affine modulus, $M_N = K_N,G_N$, such that
\begin{equation} \label{modulus1}
M = M_{A} - M_{N}.
\end{equation}
The affine modulus $M_A$ is formulated as the second derivative of the energy $E$ with respect to the homogeneous affine strain ${\epsilon}_M$ ($={\epsilon}_K,{\epsilon}_G$)~\cite{Lutsko_1989,Maloney_2004,Maloney2_2006,Maloney_2006,Lemaitre_2006,Karmakar_2010,Hentschel_2011,Zaccone_2011,Zaccone2_2011,Zaccone_2014};
\begin{equation}
\begin{aligned} \label{modulus2a}
M_{A} & = \frac{1}{V} \frac{\partial^2 E} {\partial {\epsilon_M}^{2}} = \frac{1}{V} \sum_{(i,j)} \frac{\partial^2 \phi(r_{ij})} {\partial {\epsilon_M}^{2}}, \\
&:= \frac{1}{V} \sum_{(i,j)} M_{A}^{ij}.
\end{aligned}
\end{equation}
Specifically, when we use the Green-Lagrange strain for $\epsilon_M$, then $M_A$ is formulated as the so-called Born term;
\begin{equation}
\begin{aligned} \label{modulus2}
K_{A} & = \frac{1}{V} \sum_{(i,j)} \left( \phi''(r_{ij}) - \frac{ \phi'(r_{ij}) }{ r_{ij} } \right) \frac{{ r_{ij} }^2}{9}, \\
G_{A} & = \frac{1}{V} \sum_{(i,j)} \left( \phi''(r_{ij}) - \frac{ \phi'(r_{ij}) }{ r_{ij} } \right) \frac{ {r^x_{ij}}^2 {r^y_{ij}}^2}{{r_{ij}}^2},
\end{aligned}
\end{equation}
where $r_{ij}^x, r_{ij}^y, r_{ij}^z$ are Cartesian coordinates of $\vec{r}_{ij}$; $\vec{r}_{ij}=(r_{ij}^x, r_{ij}^y, r_{ij}^z)$.
Here we note that we can also use the linear strain for $\epsilon_M$, instead of the Green-Lagrange strain~\cite{Barron_1965}.
In this case, if the stress tensor has a finite value in its components, the stress correction term is necessary in $M_A$~\cite{Lemaitre_2006,Mizuno_2013,Wittmer_2013,Wittmer2_2013,Wittmer_2015,Mizuno2_2013,Mizuno_2014,Barron_1965}, which is of same order as $\phi' \sim \Delta \varphi$.
As in Eqs.~(\ref{modulus2a}) and (\ref{modulus2}), the affine modulus $M_A$ can be decomposed into contributions from connected pairs $(i,j)$, $M_A^{ij}$, which will be shown in Sec.~\ref{sec.affine}.

On the other hand, the non-affine modulus $M_N$ is formulated in terms of the dynamical matrix $H$~\cite{Lutsko_1989,Maloney_2004,Maloney2_2006,Maloney_2006,Lemaitre_2006,Karmakar_2010,Hentschel_2011,Zaccone_2011,Zaccone2_2011,Zaccone_2014};
\begin{equation} \label{modulus3}
M_{N} = \frac{1}{V} \left( \vec{\Sigma}_M \cdot H^{-1} \cdot \vec{\Sigma}_M \right),
\end{equation}
with
\begin{equation} \label{modulus4}
\vec{\Sigma}_M = - \frac{\partial^2 E}{\partial \epsilon_M \partial \vec{r}} = - V \frac{\partial \sigma_M}{\partial \vec{r}},
\end{equation}
where $\sigma_M=(1/V) (\partial E / \partial \epsilon_M)$ is the conjugate stress to the strain $\epsilon_M$, that is the (negative) pressure $\sigma_M = -p$ for $\epsilon_M=\epsilon_K$, and the shear stress $\sigma_M=\sigma_{s}$ for $\epsilon_M=\epsilon_G$.
The pressure $p$ and the shear stress $\sigma_{s}$ are formulated through the Irving-Kirkwood expression (without the kinetic term for the static systems under study here)~\cite{Irving_1950,Allen1986};
\begin{equation}
\begin{aligned} \label{modulus5}
p &= - \frac{1}{V} \frac{\partial E}{\partial \epsilon_K} = -\frac{1}{V} \sum_{(i,j)} \phi'(r_{ij}) \frac{{ r_{ij} }}{3}, \\
\sigma_{s} &= \frac{1}{V} \frac{\partial E}{\partial \epsilon_G} = \frac{1}{V} \sum_{(i,j)} \phi'(r_{ij}) \frac{ r^x_{ij} r^y_{ij} }{r_{ij} }.
\end{aligned}
\end{equation}
Note that $\vec{\Sigma}_M=\left[ -{\partial^2 E}/{\partial \epsilon_M \partial \vec{r}_1},...,-{\partial^2 E}/{\partial \epsilon_M \partial \vec{r}_N} \right]$ is a $3N$-dimensional vector field.

Following the discussions by Maloney and Lema\^{i}tre~\cite{Maloney_2004,Maloney2_2006,Maloney_2006,Lemaitre_2006}, $\vec{\Sigma}_M$ is interpreted as the field of forces which results from an elementary affine deformation $\epsilon_M$.
This is understood when we write $\vec{\Sigma}_M$ as
\begin{equation} \label{modulus4a}
\vec{\Sigma}_M = \frac{\partial \vec{F}}{\partial \epsilon_M},
\end{equation}
where $\vec{F} = -\partial E/\partial{\vec{r}}$ is the interparticle force field acting on the $N$ particles.
In amorphous solids, $\vec{\Sigma}_M$ generally causes a force imbalance on particles, leading to an additional non-affine displacement field of the particles, $\delta \vec{R}_{\text{na} M}$ ($3N$-dimensional vector field).
Indeed, $\delta \vec{R}_{\text{na} M}$ is formulated as the linear response to the force field $\vec{\Sigma}_M$~\cite{Maloney_2004,Maloney2_2006,Maloney_2006,Lemaitre_2006};
\begin{equation} \label{modulus6}
\delta \vec{R}_{\text{na} M} = H^{-1} \cdot \vec{\Sigma}_M.
\end{equation}
From Eq.~(\ref{modulus3}), the non-affine modulus $M_N$ is the product of those two vector fields, $\vec{\Sigma}_M$ and $\delta \vec{R}_{\text{na} M}$;
\begin{equation} \label{modulus3x}
M_{N} = \frac{1}{V} \left( \vec{\Sigma}_M \cdot \delta \vec{R}_{\text{na} M} \right).
\end{equation}
Therefore $M_N$ is interpreted as an energy relaxation during the non-affine deformation, or more precisely the work done in moving the particles along the non-affine displacement field which corresponds to a repartitioning of the contact forces between particles as a result of the relaxation process.

In order to study the relation between vibrational modes $k$ and the non-affine modulus $M_N$, we formulate $M_N$ explicitly by using $\omega^k$ and $\vec{e}^k$ ($k=1,2,...,3N-3$)~\cite{Maloney_2004,Maloney2_2006,Maloney_2006,Lemaitre_2006,Karmakar_2010,Hentschel_2011,Zaccone_2011,Zaccone2_2011,Zaccone_2014}, instead of the dynamical matrix $H$.
To do this, $\vec{\Sigma}_M$ is decomposed as
\begin{equation} \label{modulus7}
\vec{\Sigma}_M = \sum_{k=1}^{3N-3} \Sigma_M^k \vec{e}^k.
\end{equation}
The component $\Sigma_M^k$ is formulated as
\begin{equation}
\begin{aligned} \label{modulus8}
\Sigma_M^k &= \vec{\Sigma}_M \cdot \vec{e}^k = -V \sum_{i=1}^{N} \frac{\partial \sigma_M}{\partial \vec{r}_{i}} \cdot \vec{e}^k_{i}, \\
&= -V \sum_{(i,j)} \frac{\partial \sigma_M}{\partial \vec{r}_{ij}} \cdot \vec{e}^k_{ij}.
\end{aligned}
\end{equation}
Here we note that the stress $\sigma_M$ is a function of $\vec{r}_{ij}$, which leads to the last equality in Eq.~(\ref{modulus8}).
Similarly $\delta \vec{R}_{\text{na} M}$ is
\begin{equation} \label{modulus7a}
\delta \vec{R}_{\text{na} M} = \sum_{k=1}^{3N-3} \delta {R}_{\text{na} M}^{k} \vec{e}^k,
\end{equation}
with
\begin{equation} \label{modulus9}
\delta {R}_{\text{na} M}^k = \delta \vec{R}_{\text{na} M} \cdot \vec{e}^k = \frac{\Sigma_M^k}{{\omega^k}^2}.
\end{equation}
The non-affine modulus $M_{N}$ can then be expressed as 
\begin{equation}
\begin{aligned} \label{modulus10}
M_{N} &= \frac{1}{V} \sum_{k=1}^{3N-3} \Sigma_M^k \delta {R}_{\text{na} M}^k= \frac{1}{V} \sum_{k=1}^{3N-3} \frac{{\Sigma_M^k}^2}{{\omega^k}^2}, \\
&:= \frac{1}{V} \sum_{k=1}^{3N-3} M_{N}^{k}.
\end{aligned}
\end{equation}
Therefore, (i) the non-affine modulus $M_N$ is decomposed into normal mode $k$ contributions, $M_{N}^{k}$, and (ii) $M_{N}^{k}$ is described as the product of the force field $\Sigma_M^k$ and the non-affine displacement field $\delta {R}_{\text{na} M}^k$, which is interpreted as an energy relaxation by the mode $k$ excitation.

In addition, from Eq.~(\ref{modulus8}), ${\Sigma_M^k}$ is interpreted as the fluctuation of the stress $\sigma_M$, induced by the mode $k$;
\begin{equation} \label{modulus8x}
\Sigma_M^k = -V \delta {\sigma}^k_M,
\end{equation}
where $\delta \sigma^k_M = \sum_{(i,j)} \left( \partial {\sigma_M}/ {\partial \vec{r}_{ij}}\right) \cdot \vec{e}^{k}_{ij}$.
Then Eq.~(\ref{modulus10}) becomes
\begin{equation} \label{modulus11}
M_{N} = \frac{1}{V} \sum_{k=1}^{3N-3} \frac{\left( V \delta {\sigma}^k_M \right)^2}{{\omega^k}^2}.
\end{equation}
Thus, (iii) the non-affine modulus $M_N$ is seen as a summation of the stress fluctuations (the pressure or shear stress fluctuations).
In fact, at finite temperatures $T$, the non-affine modulus is formulated in terms of thermal fluctuations of the stress~\cite{Lutsko_1989,Mayr_2009,Mizuno_2013,Wittmer_2013,Wittmer2_2013,Wittmer_2015,yoshimoto_2004,Zaccone_2013,Mizuno2_2013,Mizuno_2014}.
Eqs.~(\ref{modulus10}) and~(\ref{modulus11}) allow us to directly relate the vibrational normal modes $k$ to the non-affine modulus $M_N$, which will be done in Secs.~\ref{sec.nonaffine1} and~\ref{sec.nonaffine2}.

\begin{figure}[t]
\centering
\includegraphics[width=0.48\textwidth]{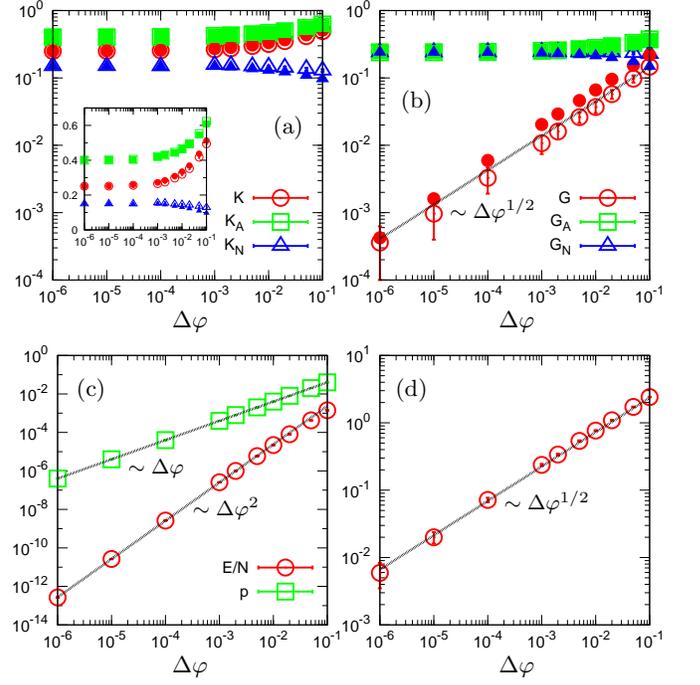}
\vspace*{0mm}
\caption{\label{fig.phidependence}
(Color online) Dependence on the packing fraction, $\Delta \varphi = \varphi-\varphi_c$, of the (a) bulk modulus $K, K_A, K_N$, (b) shear modulus $G, G_A, G_N$, (c) potential energy per particle $E/N$, pressure $p$, and the (d) excess contact number $\Delta z=z-z_c$.
In (a) and (b), we plot values from the unstressed system (closed symbols), in addition to values from the original stressed system (open symbols).
The inset to (a) presents $K, K_A, K_N$ on a linear scale.
The lines indicate power-law scalings with respect to $\Delta \varphi$.
The error bars were calculated from $100$ configuration realizations.}
\end{figure}

\section{Results} \label{results}
\subsection{Dependence of elastic moduli on packing fraction $\Delta \varphi$}
{\bf Scaling laws with packing fraction $\boldsymbol{\Delta \varphi}$.}
Figure~\ref{fig.phidependence} shows the elastic moduli $K, G$, potential energy per particle $E/N$, pressure $p$, and the excess contact number $\Delta z=z-z_c$, as functions of $\Delta \varphi$.
Our values as well as the power-law scalings are consistent with previous works on the harmonic system~\cite{OHern_2002,OHern_2003};
\begin{equation}
\begin{aligned} \label{plmacro}
& K \sim \Delta \varphi^0, \qquad G \sim \Delta \varphi^{1/2}, \\
& E \sim \Delta \varphi^2, \qquad p \sim \Delta \varphi, \\
& \Delta z \sim \Delta \varphi^{1/2}.
\end{aligned}
\end{equation}
As $\Delta \varphi \rightarrow 0$, the affine shear modulus $G_A$ and the non-affine shear modulus $G_N$ converge to the same value, and consequently the total shear modulus $G$ vanishes according to $G \sim \Delta \varphi^{1/2} \rightarrow 0$.
On the other hand, the affine bulk modulus $K_A$ is always larger than the non-affine value $K_N$, i.e., $K_A > K_N$, and the total bulk modulus $K$ does not vanish, approaching a finite constant value.

{\bf Comparison between stressed and unstressed systems.}
The stressed and unstressed systems show similar values of $K$ and $G$, as well as consistent exponents for the power-law scalings (compare open and closed symbols in Fig.~\ref{fig.phidependence}(a),(b)).
Close to the transition point ($\Delta \varphi \ll 1$), the interparticle force, $\sim \phi'(r_{ij}) \sim \mathcal{O}(\Delta \varphi)$, becomes very small, as manifested in the pressure, $p \sim \phi'(r_{ij}) \sim \Delta \varphi \ll 1$.
In this situation, the unstressed system is a good approximation to the original stressed system~\cite{Wyart_2005,Wyart_2006,Xu_2007}.
However, as we will see in Figs.~\ref{fig.decomposition} and~\ref{fig.decompositiona} and discuss in Sec.~\ref{sec.nonaffine1}, differences between the two systems visibly appear in the non-affine modulus contributions, $M_{N}^{k}$, from the low-$\omega$ normal modes $k$.
These differences are hidden by a summation of $M_N^k$ over all $3N-3$ normal modes, and as a result, only tiny differences are noticeable in the total moduli, $K$ and $G$ (or $K_N$ and $G_N$) (Fig.~\ref{fig.phidependence}(a),(b)).

\begin{figure}[t]
\centering
\includegraphics[width=0.48\textwidth]{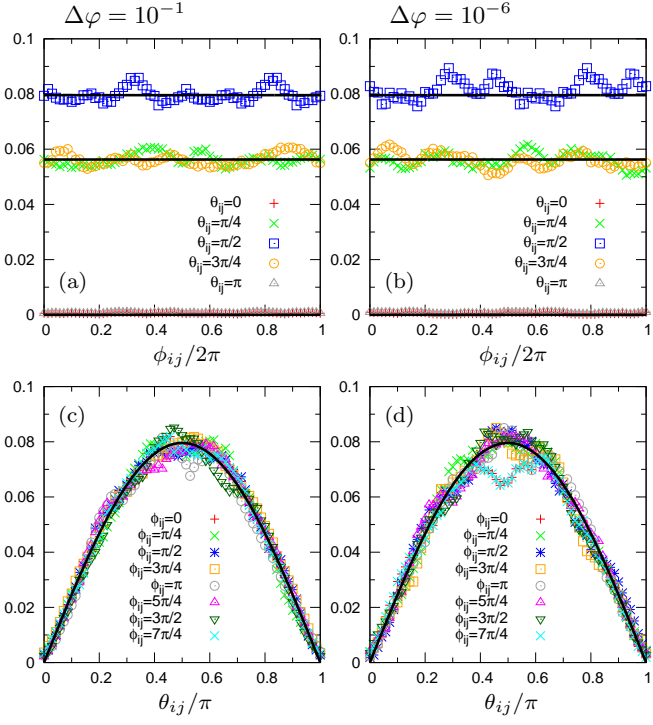}
\vspace*{0mm}
\caption{\label{fig.angle}
(Color online) Probability distribution, $P(\phi_{ij},\theta_{ij})$, of the orientation angles of the unit bond vector, $\vec{n}_{ij}=\left(\cos\phi_{ij} \sin \theta_{ij},\sin\phi_{ij} \sin \theta_{ij},\cos \theta_{ij} \right)$.
We plot $P(\phi_{ij},\theta_{ij})$ as a function of $\phi_{ij}$ in (a),(b), and $\theta_{ij}$ in (c),(d).
Note $0\le \phi_{ij} < 2\pi$, and $0\le \theta_{ij}\le \pi$, and in the figures, $\phi_{ij}$ and $\theta_{ij}$ are normalized by $2\pi$ and $\pi$, respectively.
The packing fraction is $\Delta \varphi = 10^{-1}$ in left panels and $10^{-6}$ in right panels.
The solid lines indicate $P(\phi_{ij},\theta_{ij})$ in Eq.~(\ref{affine2}), which coincides with numerical results (symbols), thereby demonstrating the isotropic distribution of the orientation of $\vec{n}_{ij}$.}
\end{figure}

\begin{figure}[t]
\centering
\includegraphics[width=0.48\textwidth]{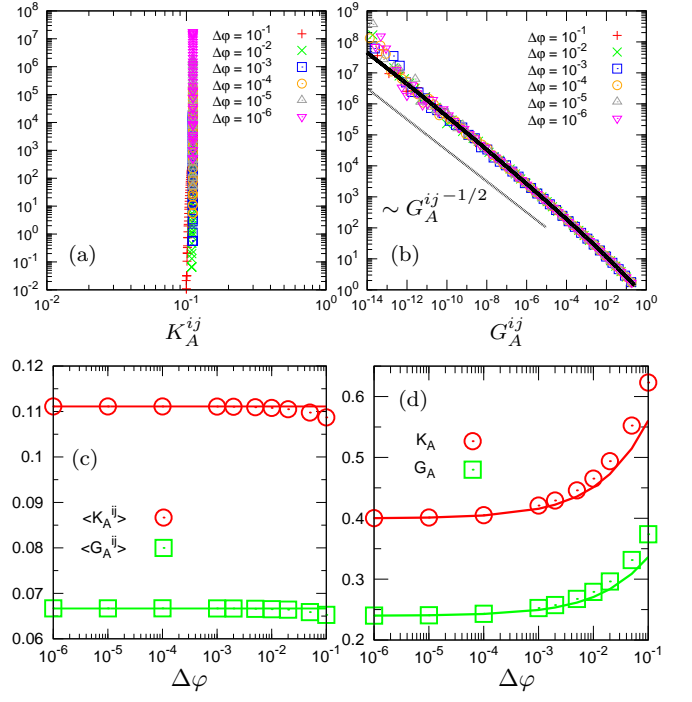}
\vspace*{0mm}
\caption{\label{fig.affine}
(Color online) Contributions to the affine moduli, $K_A^{ij}$ and $G_A^{ij}$, from each connected pair of particles, $(i,j)$ (see Eqs.~(\ref{modulus2a}) and~(\ref{modulus2})).
Shown are the probability distributions, (a) $P(K^{ij}_A)$ and (b) $P(G^{ij}_A)$, for the range of $10^{-6} \le \Delta \varphi \le 10^{-1}$.
It is seen that $P(K^{ij}_A) \simeq \delta (K^{ij}_A -1/9)$ (delta function), and $P(G^{ij}_A) \sim {G^{ij}_A}^{-1/2}$ (power-law function).
The solid line in (b) presents $P(G_A^{ij})$ calculated from Eq.~(\ref{affine2a}).
In (c), the average values over all $N^\text{ct}$ contacts, $\left< K^{ij}_A \right>$ and $\left< G^{ij}_A \right>$, are plotted as functions of $\Delta \varphi$.
The horizontal lines indicate the values of $\left< K^{ij}_A \right>=1/9$ and $\left< G^{ij}_A \right>=1/15$ (see Eq.~(\ref{affine3})).
In (d), we compare $K_A$ and $G_A$ from Eq.~(\ref{affine5}) (lines) to numerical values presented in Fig.~\ref{fig.phidependence}(a),(b) (symbols).}
\end{figure}

\subsection{Affine moduli} \label{sec.affine}
Firstly we study the affine modulus $M_A$, which is decomposed into contributions from each contact $(i,j)$, $M_{A}^{ij}$, as in Eqs.~(\ref{modulus2a}) and~(\ref{modulus2}).
Close to the transition point $\varphi_c$, $r_{ij} = 1 + \mathcal{O}(\Delta \varphi)$, $\phi'(r_{ij}) = \mathcal{O}(\Delta \varphi)$, and $\phi''(r_{ij}) = 1$ for all contacts, $(i,j) \in N^\text{ct}$.
Therefore, we get
\begin{equation}
\begin{aligned} \label{affine1}
K_{A}^{ij} &= \left( \phi''(r_{ij}) - \frac{ \phi'(r_{ij}) }{ r_{ij} } \right) \frac{ { r_{ij} }^2}{9},\\
&= \frac{1}{9} + \mathcal{O}(\Delta \varphi),\\
G_{A}^{ij} &= \left( \phi''(r_{ij}) - \frac{ \phi'(r_{ij}) }{ r_{ij} } \right) \frac{ {r^x_{ij}}^2 {r^y_{ij}}^2}{{r_{ij}}^2},\\
&= {n_{ij}^{x}}^2 {n_{ij}^{y}}^2 + \mathcal{O} (\Delta \varphi),\\
&= \cos^2\phi_{ij} \sin^2 \phi_{ij} \sin^4 \theta_{ij} + \mathcal{O} (\Delta \varphi).
\end{aligned}
\end{equation}
In the last equality for $G_A^{ij}$ of Eq.~(\ref{affine1}), we write the unit bond vector, $\vec{n}_{ij}= \left( n_{ij}^x,n_{ij}^y,n_{ij}^z \right)$, as
\begin{equation}
\left( n_{ij}^x,n_{ij}^y,n_{ij}^z \right) = \left(\cos\phi_{ij} \sin \theta_{ij},\sin\phi_{ij} \sin \theta_{ij},\cos \theta_{ij} \right),
\end{equation}
where the pair of angles, $(\phi_{ij},\ \theta_{ij})$, are the polar coordinates specifying the orientation of $\vec{n}_{ij}$, and $0 \le \phi_{ij} < 2\pi$, $0 \le \theta_{ij}\le \pi$.
The bulk modulus, $K_A^{ij} \simeq 1/9$ ($=\phi''(r_{ij})/9$), just picks up the stiffness of bond $\vec{n}_{ij}$, which is same for all contacts.
Whereas the shear modulus, $G_A^{ij} \simeq {n_{ij}^{x}}^2 {n_{ij}^{y}}^2 =\cos^2\phi_{ij} \sin^2 \phi_{ij} \sin^4 \theta_{ij}$, depends on the orientation of $\vec{n}_{ij}$.
In the present work, we follow Zaccone \textit{et. al.}~\cite{Zaccone_2011,Zaccone2_2011} and assume an isotropic distribution of the orientation of $\vec{n}_{ij}$: The joint probability distribution of $\phi_{ij},\theta_{ij}$ is assumed to be
\begin{equation} \label{affine2}
P(\phi_{ij},\theta_{ij}) = \frac{1}{2\pi} \times \frac{\sin \theta_{ij}}{2}.
\end{equation}
We plot numerical results of $P(\phi_{ij},\theta_{ij})$ for the packing fractions of high $\Delta \varphi = 10^{-1}$ and low $\Delta \varphi = 10^{-6}$ in Fig.~\ref{fig.angle}, which well verifies Eq.~(\ref{affine2}).

{\bf Probability distribution of $\boldsymbol{M_A^{ij}}$.}
Figure~\ref{fig.affine} presents the probability distributions, $P(K_A^{ij})$ in (a) and $P(G_A^{ij})$ in (b).
We see that $P(K_A^{ij})$ and $P(G_A^{ij})$ are both insensitive to $\Delta \varphi$.
As expected from Eq.~(\ref{affine1}), $P(K_A^{ij})$ shows a delta function, $P(K_A^{ij}) \simeq \delta(K_A^{ij}-1/9)$.
On the other hand, $P(G_A^{ij})$ is a power-law function, $P(G_A^{ij}) \sim {G_A^{ij}}^{-1/2}$, with a finite range of $0 \le G^{ij}_A \le 1/4$.
The power-law  behavior of $P(G_A^{ij})$ is obtained using the isotropic distribution of the bond-orientation, i.e., $P(\phi_{ij},\theta_{ij})$ in Eq.~(\ref{affine2}), as
\begin{equation}
\begin{aligned} \label{affine2a}
& P(G_A^{ij}) dG_A^{ij} \\
& = \int_{G_A^{ij} < \cos^2\phi_{ij} \sin^2 \phi_{ij} \sin^4 \theta_{ij} < G_A^{ij}+dG_A^{ij}} P(\phi_{ij},\theta_{ij}) d\phi_{ij} d\theta_{ij}, \\
& \Longrightarrow\\
& P(G_A^{ij}) \\
&=\frac{{G_A^{ij}}^{-1/2}}{\pi} \int_{2{G^{ij}_A}^{1/2}}^{1} \left[ x(1-x^2) \left(x- 2 {G_A^{ij}}^{1/2} \right) \right]^{-1/2} dx.
\end{aligned}
\end{equation}
We note that $G^{ij}_A$ takes values in the range of $0 \le G^{ij}_A \le 1/4 \Leftrightarrow 0 \le 2{G^{ij}_A}^{1/2} \le 1$.
Eq.~(\ref{affine2a}) is numerically verified in Fig.~\ref{fig.affine}(b) (see solid line), and demonstrates that the power-law  behavior, $P(G_A^{ij}) \sim {G_A^{ij}}^{-1/2}$, comes from its prefactor.

{\bf Average value $\boldsymbol{\left< M_A^{ij} \right>}$.}
From the distribution function $P(M^{ij}_A)$, we obtain the average value $\left< M_A^{ij} \right>$;
\begin{equation}
\begin{aligned} \label{affine3}
\left< K_A^{ij} \right> &= \frac{1}{N^\text{ct}} \sum_{(i,j) \in N^\text{ct}} K_A^{ij} = \int K_A^{ij} P(K_A^{ij}) dK_A^{ij}, \\
&= \frac{1}{9} + \mathcal{O} (\Delta \varphi),\\
\left< G_A^{ij} \right> &= \frac{1}{N^\text{ct}} \sum_{(i,j) \in N^\text{ct}} G_A^{ij} = \int G_A^{ij} P(G_A^{ij}) dG_A^{ij}, \\ 
&= \frac{1}{15} + \mathcal{O} (\Delta \varphi),
\end{aligned}
\end{equation}
where $\left< \right>$ denotes the average over all the $N^\text{ct}$ contacts, $(i,j)$.
$\left< G_A^{ij} \right> ={1}/{15}$ can be also calculated by using $P(\phi_{ij},\theta_{ij})$ in Eq.~(\ref{affine2}) as
\begin{equation}
\begin{aligned}
\left< G_A^{ij} \right> &= \int_{0}^{2\pi} d\phi_{ij} \int_0^\pi d\theta_{ij} P(\phi_{ij},\theta_{ij}) G_A^{ij},\\
&= \int_{0}^{2\pi} \frac{d\phi_{ij}}{2\pi} \int_0^\pi \frac{\sin \theta_{ij} d\theta_{ij}}{2} \cos^2\phi_{ij} \sin^2 \phi_{ij} \sin^4 \theta_{ij},\\
&= \frac{1}{15}.
\end{aligned}
\end{equation}
Panel (c) of Fig.~\ref{fig.affine} plots numerical values of $\left< M_A^{ij} \right>$ as a function of $\Delta \varphi$, and verifies Eq.~(\ref{affine3}).

{\bf Formulation of the affine modulus $\boldsymbol{M_A}$.}
The total affine modulus $M_{A}$ is therefore formulated as
\begin{equation}
\begin{aligned} \label{affine5x}
M_{A} &= \frac{1}{V} \left< M_{A}^{ij} \right> N^\text{ct} = \frac{\hat{\rho}}{2} \left< M_{A}^{ij} \right> (z_c + \Delta z),\\
&= M_{Ac} + \frac{\hat{\rho}_c}{2} \left< M_{A}^{ij} \right>_c \Delta z + \mathcal{O} (\Delta \varphi),
\end{aligned}
\end{equation}
where $M_{Ac} =  \left({\hat{\rho}_c}/{2}\right) \left< M_{A}^{ij} \right>_c z_c $ is the critical value at the transition point $\varphi_c$.
Specifically, we get
\begin{equation}
\begin{aligned} \label{affine5}
K_{A} &= K_{Ac} + \frac{\hat{\rho}_c}{18} \Delta z + \mathcal{O}(\Delta \varphi),\\
G_{A} &= G_{Ac} + \frac{\hat{\rho}_c}{30} \Delta z + \mathcal{O}(\Delta \varphi),
\end{aligned}
\end{equation}
with
\begin{equation}
\begin{aligned} \label{affine5a}
K_{Ac} &= \frac{\hat{\rho}_c}{18}z_c \simeq 0.40, \\
G_{Ac} &= \frac{\hat{\rho}_c}{30}z_c \simeq 0.24.
\end{aligned}
\end{equation}
We note that $\Delta z \sim \Delta \varphi^{1/2}$ is the leading order term of $M_A$ in Eqs.~(\ref{affine5x}) and~(\ref{affine5}).
Eq.~(\ref{affine5}) is the same formulation obtained by Zaccone \textit{et. al.}~\cite{Zaccone_2011,Zaccone2_2011} for $d=3$ dimensions, which is based on the isotropic distribution of the bond-orientations, $P(\phi_{ij},\theta_{ij})$ in Eq.~(\ref{affine2}).
Figure~\ref{fig.affine}(d) demonstrates that Eq.~(\ref{affine5}) matches the numerical values of $M_A$ presented in Fig.~\ref{fig.phidependence}(a),(b).
On approach to the transition point $\varphi_c$, the excess contact number $\Delta z$ is vanishing, which reduces the affine modulus $M_A$ towards the critical value $M_{Ac}$.
It is worth mentioning that the critical values of both $K_{Ac}$ and $G_{Ac}$ are finite positive (see Eq.~(\ref{affine5a})).
Therefore, similar to the coordination number $z$, $M_A$ discontinuously drops to zero, through the transition to the fluid phase, $\varphi < \varphi_c$, where $M_A \equiv 0$.

\begin{figure}[t]
\centering
\includegraphics[width=0.48\textwidth]{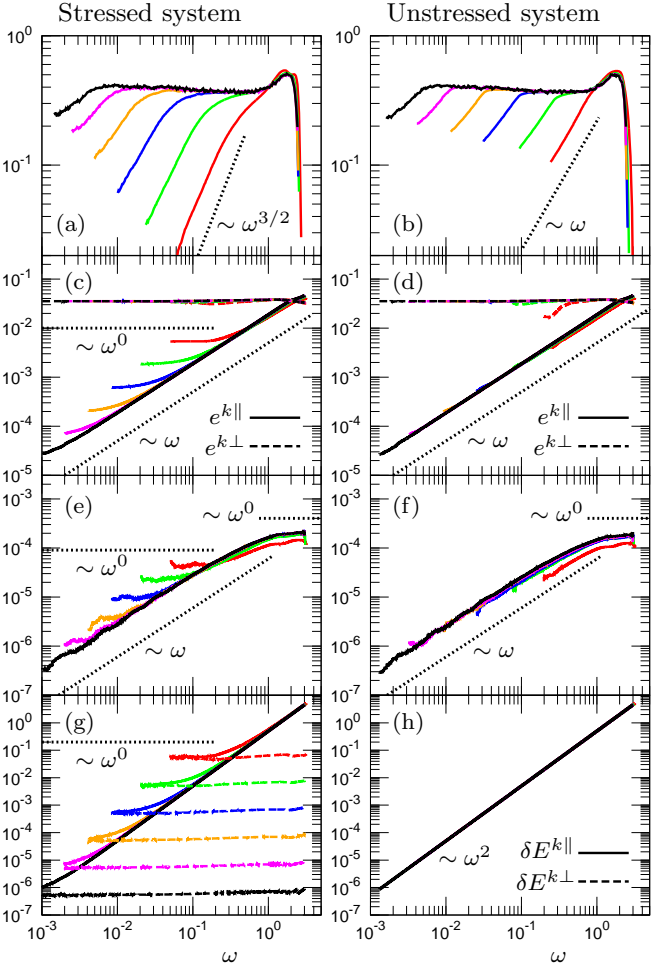}
\vspace*{0mm}
\caption{\label{fig.vibration}
(Color online) Vibrational eigenmodes in the stressed (left panels) and the unstressed (right panels) systems.
The vDOS $g(\omega)$ in (a),(b), displacements ${e}^{k \parallel}$ (solid lines), ${e}^{k \perp}$ (dashed lines) in (c),(d), net displacement ${e}^{k \parallel}_\text{net}$ in (e),(f), and the mode energies $\delta E^{k \parallel}$ (solid), $\delta E^{k \perp}$ (dashed) in (g),(h), are plotted as functions of the eigenfrequency $\omega$.
See Eqs.~(\ref{displace1}) and~(\ref{displace2}) for the definitions of ${e}^{k \parallel}, {e}^{k \perp}, {e}^{k \parallel}_\text{net}$.
The values of ${e}^{k \parallel}, {e}^{k \perp}, {e}^{k \parallel}_\text{net}, \delta E^{k \parallel}, \delta E^{k \perp}$ are averaged over frequency bins of $\log_{10} \omega^k \in [\log_{10} \omega-\Delta \omega/2,\log_{10} \omega + \Delta \omega/2]$ with $\Delta \omega = 0.07$.
The different lines indicate different packing fractions, $\Delta \varphi = 10^{-1}$ (red), $10^{-2}$ (green), $10^{-3}$ (blue), $10^{-4}$ (orange), $10^{-5}$ (magenta), $10^{-6}$ (black), from right to left or from top to bottom.
Details of the presented quantities are given in Sec.~\ref{sec.vibration}.}
\end{figure}

\begin{figure}[t]
\centering
\includegraphics[width=0.48\textwidth]{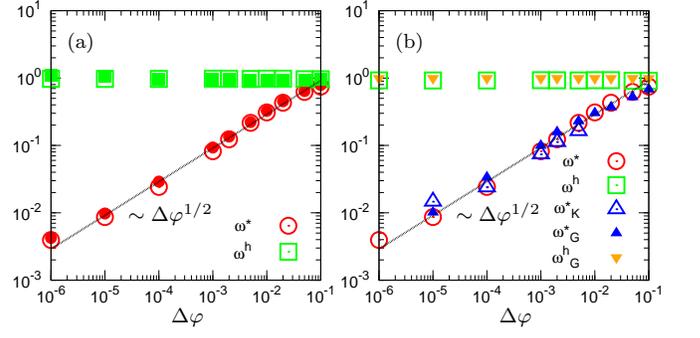}
\vspace*{0mm}
\caption{\label{fig.frequency}
(Color online) Characteristic frequencies, $\omega^\ast$, $\omega^h$, $\omega^\ast_{M}$, and $\omega_{M}^h$ ($M=K, G$), as functions of $\Delta \varphi$.
$\omega^\ast, \omega^h$ characterize the vDOS $g(\omega)$, and $\omega^\ast_{M}, \omega^h_{M}$ are from the modal contribution to the non-affine moduli, $M_N^k = K_N^k, G_N^k$.
In (a), we compare $\omega^\ast$, $\omega^h$ between the stressed (open symbols) and unstressed (closed symbols) systems, which are seen to coincide with each other.
In (b), $\omega^\ast$, $\omega^h$ are compared to $\omega^\ast_{M}$, $\omega_{M}^h$ for the stressed system.
We observe that $\omega^\ast \simeq \omega^\ast_{M} \sim \Delta \varphi^{1/2}$ whereas $\omega^h \simeq \omega^h_{M} \simeq 1.0$ is insensitive to $\Delta \varphi$.
Note that for the bulk modulus $M=K$, only $\omega^\ast_{K}$ is determined in $\Delta \varphi \le 5 \times 10^{-3}$ ($\omega_{K}^h$ is not).
A more detailed discussion of these frequencies is given in the main text.}
\end{figure}

\begin{figure}[t]
\centering
\includegraphics[width=0.48\textwidth]{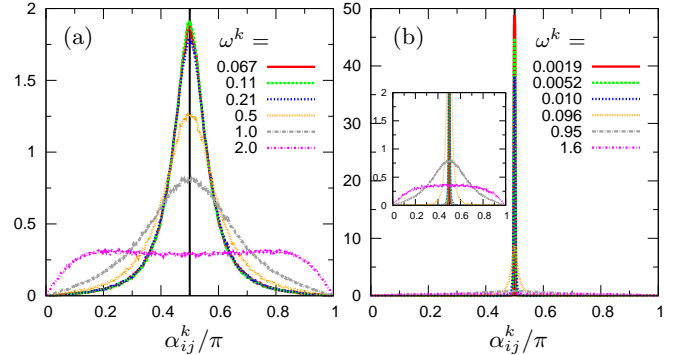}
\vspace*{0mm}
\caption{\label{fig.alpha}
(Color online) Probability distribution $P(\alpha_{ij}^k)$ of the sliding angle, $\alpha^k_{ij} = \arctan \left( \left| \vec{e}_{ij}^{k \perp} \right| / \left| \vec{e}_{ij}^{k \parallel} \right| \right)$, for several different vibrational modes $k$, at (a) $\Delta \varphi=10^{-1}$ and (b) $\Delta \varphi=10^{-6}$ (inset is a zoom of the central portion).
The number of the label indicates the eigenfrequency $\omega^k$.
The value of $\alpha^k_{ij}$ is normalized by $\pi$, and the vertical solid line indicates $\alpha^k_{ij}=\pi/2$.}
\end{figure}

\subsection{Vibrational eigenmodes} \label{sec.vibrationalstate}
Before studying the non-affine modulus $M_N$, we report on the vibrational eigenmodes in this section.
As explained in Sec.~\ref{sec.vibration}, we characterize vibrational mode $k$ in terms of its eigenfrequency $\omega^k$, eigenvectors $\vec{e}_{ij}^{k \parallel}, \vec{e}_{ij}^{k \perp}$, and mode energies $\delta E^{k \parallel}, \delta E^{k \perp}$.
Regarding the eigenvectors $\vec{e}_{ij}^{k \parallel}, \vec{e}_{ij}^{k \perp}$ (see Eq.~(\ref{vs2})), we introduce the ``absolute'' displacement ${e}^{k \parallel}, {e}^{k \perp}$ (root mean square);
\begin{equation}
\begin{aligned} \label{displace1}
{e}^{k \parallel} &= \sqrt{\frac{1}{N^\text{ct}} \sum_{(i,j) \in N^\text{ct}} {\vec{e}^{k \parallel}_{ij}}^2 } = \sqrt{ \left< {\vec{e}^{k \parallel}_{ij}}^2 \right>}, \\
{e}^{k \perp} &= \sqrt{\frac{1}{N^\text{ct}} \sum_{(i,j) \in N^\text{ct}} {\vec{e}^{k \perp}_{ij}}^2} = \sqrt{ \left< {\vec{e}^{k \perp}_{ij}}^2 \right>},
\end{aligned}
\end{equation}
and the ``net'' displacement ${e}^{k \parallel}_\text{net}$;
\begin{equation} \label{displace2} 
{e}^{k \parallel}_\text{net} = \left| \frac{1}{N^\text{ct}} \sum_{(i,j) \in N^\text{ct}} \vec{e}^{k \parallel}_{ij} \cdot \vec{n}_{ij} \right| = \left| \left< \vec{e}^{k \parallel}_{ij} \cdot \vec{n}_{ij} \right> \right|.
\end{equation}
In this way, the net displacement ${e}^{k \parallel}_\text{net}$ is a measure of vibrational motions $\vec{e}_{ij}^{k \parallel}$ along the bond vector $\vec{n}_{ij}$ that distinguishes between compressing ($\vec{e}^{k \parallel}_{ij} \cdot \vec{n}_{ij}<0$) and stretching ($\vec{e}^{k \parallel}_{ij} \cdot \vec{n}_{ij}>0$) motions, while ${e}^{k \parallel}$ merely picks up the ``absolute'' amplitude.
The absolute amplitudes of ${e}^{k \parallel}, {e}^{k \perp}$ are directly related to the energies $\delta E^{k \parallel}$ and $\delta E^{k \perp}$ (see Eq.~(\ref{vs1}));
\begin{equation}
\begin{aligned} \label{displace3}
\delta E^{k \parallel} & = N^\text{ct} \left< \frac{\phi''(r_{ij})}{2} {\vec{e}^{k \parallel}_{ij}}^2 \right> \sim {{e}^{k \parallel}}^2,\\
\delta E^{k \perp} & = N^\text{ct} \left< -\frac{\phi'(r_{ij})}{2 r_{ij}} {\vec{e}^{k \perp}_{ij}}^2 \right> \sim \Delta \varphi {{e}^{k \perp}}^2,
\end{aligned}
\end{equation}
whereas the net amplitude of ${e}^{k \parallel}_\text{net}$ is related to the force $\left| \Sigma_M^k \right|$ and the non-affine displacement $\left| \delta {R}_{\text{na} M}^k \right|$ (see Eqs.~(\ref{modulus8}) and~(\ref{modulus9}));
\begin{equation}
\begin{aligned} \label{displace4}
& \left| \Sigma_M^k \right| \sim \left| N^\text{ct} \left< \phi''(r_{ij}) \left( \vec{e}^{k \parallel}_{ij}\cdot \vec{n}_{ij} \right) \right> + \mathcal{O}(\Delta \varphi) \right| \sim {e}^{k \parallel}_\text{net}, \\
& \left| \delta {R}_{\text{na} M}^k \right| = \frac{ \left| \Sigma_M^k \right| }{ {\omega}^{2} } \sim \frac{ {e}^{k \parallel}_\text{net} }{ {\omega}^{2} }.
\end{aligned}
\end{equation}

Figure~\ref{fig.vibration} shows $g(\omega)$ (vDOS), ${e}^{k \parallel}, {e}^{k \perp}, {e}^{k \parallel}_\text{net}, \delta E^{k \parallel}, \delta E^{k \perp}$ as functions of the eigenfrequency $\omega$, for the range of $\Delta \varphi = 10^{-1}$ to $10^{-6}$.
In the figure, the values of ${e}^{k \parallel}, {e}^{k \perp}, {e}^{k \parallel}_\text{net}, \delta E^{k \parallel}, \delta E^{k \perp}$ are averaged over frequency bins of $\log_{10} \omega^k \in [\log_{10} \omega-\Delta \omega/2,\log_{10} \omega + \Delta \omega/2]$ with $\Delta \omega = 0.07$.
Results from the original stressed system (left panels) as well as the unstressed system (right panels) are presented.

{\bf Vibrational density of states $\boldsymbol{g(\omega)}$.}
As reported in previous studies~\cite{Silbert_2009,Xu_2010,Silbert_2005}, the vDOS $g(\omega)$, presented in Fig.~\ref{fig.vibration}(a),(b), is divided into three regimes distinguishable by two characteristic frequencies $\omega^\ast$ and $\omega^h$; (i) intermediate $\omega^\ast< \omega < \omega^h$ regime, (ii) low $\omega < \omega^\ast$ regime, and (iii) high $\omega > \omega^h$ regime.
Over the intermediate regime, $\omega^\ast< \omega < \omega^h$, $g(\omega)$ is nearly constant, i.e., $g(\omega)$ exhibits a plateau.
At the low-frequency end, $\omega < \omega^\ast$, $g(\omega)$ decreases to zero as $\omega \rightarrow 0$, following Debye-like, power-law behavior, $g(\omega) \sim \omega^{a}$.
Although, here we find the values of the exponents, $a \simeq 3/2$ in the stressed system and $a \simeq 1$ in the unstressed system, which are both smaller than the exact Debye exponent, $a = d-1 = 2$~\cite{Ashcroft,kettel,McGaughey}.
(We would expect to recover the Debye behavior, $g(\omega) \sim \omega^{2}$, in the low frequency limit.)
Finally, at the high $\omega > \omega^h$, $g(\omega)$ goes to zero as $\omega$ increases to $\omega_\text{max} \simeq 3$, where the vibrational modes are highly localized~\cite{Silbert_2009,Xu_2010}.
In Fig.~\ref{fig.frequency}(a), we show the characteristic frequencies, $\omega^\ast$ and $\omega^h$, as functions of $\Delta \varphi$.
As $\Delta \varphi \rightarrow 0$, $\omega^\ast$ goes to zero, following the power-law scaling of $\omega^\ast \sim \Delta \varphi^{1/2} \rightarrow 0$~\cite{Silbert_2005,Wyart_2005,Wyart_2006}, whereas $\omega^h \simeq 1.0$ is almost constant, independent of $\Delta \varphi$, and is set by the particle stiffness (recall, $\textrm{k}=1.0$).
Thus, as demonstrated in Figs.~\ref{fig.vibration}(a),(b) and~\ref{fig.frequency}(a), on approach to the transition point $\varphi_c$, (i) the intermediate plateau regime extends towards zero frequency, (ii) the low $\omega <\omega^\ast$ region shrinks and disappears, and (iii) the high $\omega >\omega^h$ regime remains unchanged.

Figure~\ref{fig.frequency}(a) also compares $\omega^\ast, \omega^h$ between the stressed (open symbols) and the unstressed (closed symbols) systems, and demonstrates that the two systems show identical values of $\omega^\ast, \omega^h$.
Thus, the three regimes, (i) to (iii), in $g(\omega)$ practically coincide between the two systems.
However, here we note that the crossover at $\omega = \omega^\ast$ between regimes (i) and (ii) is milder in the stressed system than in the unstressed system, which is clearly observed in Fig~\ref{fig.vibration}(a),(b) and was reported in previous works~\cite{Wyart_2005,Wyart_2006,Xu_2007}.
The stress, $\sim \phi'(r_{ij})$, reduces the mode energy $\delta E^k$ by $\delta E^{k \perp}$ (see Eq.~(\ref{vs1})), and shifts the vibrational modes to the low $\omega$ side~\cite{Ellenbroek_2006,Ellenbroek_2009,Ellenbroek2_2009,Wyart_2005,Wyart_2006,Xu_2007}.
Thus, the ``anomalous modes", which lie in the plateau regime, move into the Debye-like regime, and as a result, the crossover becomes less abrupt in the stressed system.

{\bf Displacements $\boldsymbol{{e}^{k \parallel}, {e}^{k \perp}}$.}
We now pay attention to the stressed system in the left panels of Fig.~\ref{fig.vibration}.
When looking at ${e}^{k \parallel}$ (solid lines) and ${e}^{k \perp}$ (dashed lines) in (c), the sliding displacement ${e}^{k \perp}$ is almost constant, i.e., ${e}^{k \perp} \simeq {A}^{\perp}$.
In the tangential direction, particles are displaced by the same magnitude in each mode $k$, independent of the eigenfrequency $\omega^k$.
Since there are few constraints in the tangential direction close to the jamming transition, the sliding motion ${e}^{k \perp}$ dominates over the normal motion ${e}^{k \parallel}$ and determines the whole vibrational motion regardless of the mode frequency $\omega^k$ (except for the highest frequency end).

On the other hand, the compressing/stretching displacement ${e}^{k \parallel}$ is comparable to ${e}^{k \perp}$ at high $\omega$, and as $\omega$ is lowered, it monotonically decreases, following ${e}^{k \parallel} \sim \omega$.
Around $\omega = \omega^\ast$, ${e}^{k \parallel}$ shows a functional crossover, from ${e}^{k \parallel} \sim \omega$ to $\sim \omega^0$.
As $\omega \rightarrow 0$, ${e}^{k \parallel}$ converges to a constant value, ${A}^{\parallel}$, which depends on $\Delta \varphi$; ${e}^{k \parallel} \rightarrow {A}^{\parallel}(\Delta \varphi)$.
Here we note that as $\omega$ decreases, ${e}^{k \perp}$ increases relative to ${e}^{k \parallel}$, indicating that the sliding angle, $\alpha^k_{ij} := \arctan \left(\left| \vec{e}_{ij}^{k \perp} \right| / \left| \vec{e}_{ij}^{k \parallel} \right| \right)$, approaches $\pi/2$ for each contact $(i,j)$, and vibrational motions become more floppy-like \cite{Ellenbroek_2006,Ellenbroek_2009,Ellenbroek2_2009}.
To illustrate this point more explicitly, Fig.~\ref{fig.alpha} plots the probability distribution $P(\alpha_{ij}^k)$ for several different normal modes $k$, and shows that the lower $\omega^k$ mode expresses a higher probability for $\alpha_{ij}^k = \pi/2$.
At low packing fraction $\Delta \varphi=10^{-6}$ (Fig.~\ref{fig.alpha}(b)), the lowest $\omega^k$ modes resemble a delta function distribution, $P(\alpha_{ij}^k) \simeq \delta(\alpha_{ij}^k-\pi/2)$, where sliding ${e}^{k \perp}$ is orders of magnitude larger than compressing/stretching ${e}^{k \parallel}$.

{\bf Mode energies $\boldsymbol{\delta E^{k \parallel}, \delta E^{k \perp}}$.}
We next turn to the mode energies, $\delta E^{k \parallel}$ (solid lines) and $\delta E^{k \perp}$ (dashed lines), in Fig.~\ref{fig.vibration}(g).
From Eq.~(\ref{displace3}) and ${e}^{k \perp} \simeq A^{\perp}$, the transverse energy $\delta E^{k \perp}$ is described as $\delta E^{k \perp} \sim \Delta \varphi {A^{\perp}}^2 \sim \Delta \varphi$.
Thus, $\delta E^{k \perp}$ is independent of $\omega$ and is proportional to $\Delta \varphi$, which is indeed numerically demonstrated in (g).

On the other hand, the compressing/stretching energy $\delta E^{k \parallel}$ dominates over $\delta E^{k \perp}$ at high $\omega$, and the total mode energy is determined by $\delta E^{k \parallel}$ only; $\delta E^{k \parallel} \simeq \delta E^{k}$.
As $\omega$ is lowered, $\delta E^{k \parallel}$ decreases as $\delta E^{k \parallel} \simeq \delta E^{k} = {\omega}^2/2$ (see Eq.~(\ref{vs3})).
From Eq.~(\ref{displace3}) we obtain $\delta E^{k \parallel} \sim {{e}^{k \parallel}}^2 \sim {\omega}^2$, which explains the behavior of ${e}^{k \parallel} \sim \omega$ in (c).
At the crossover $\omega = \omega^\ast$, $\delta E^{k \parallel} \simeq {{\omega}^\ast}^2/2$ reaches the same order of magnitude as $\delta E^{k \perp}$, from which we obtain the scaling law of $\omega^\ast$ with respect to $\Delta \varphi$ as
\begin{equation} \label{resultvs1}
\begin{aligned}
& \delta E^{k \parallel} \simeq \frac{{{\omega}^\ast}^2}{2} \sim \delta E^{k \perp} \sim \Delta \varphi, \\
& \Longleftrightarrow \ {\omega}^\ast \sim {\delta E^{k \perp}}^{1/2} \sim \Delta \varphi^{1/2}.
\end{aligned}
\end{equation}
Eq.~(\ref{resultvs1}) is indeed what we observed in Fig.~\ref{fig.frequency} and is consistent with previous works~\cite{Silbert_2005,Wyart_2005,Wyart_2006}.
The crossover in ${e}^{k \parallel}$ at $\omega=\omega^\ast$ corresponds to that in $\delta E^{k \parallel}$.
As $\omega$ further decreases towards zero frequency, $\delta E^{k \parallel}$ converges to $\delta E^{k \perp} \sim \Delta \varphi$ such that the total $\delta E^k = \delta E^{k \parallel} - \delta E^{k \perp} \rightarrow 0$, thus ${e}^{k \parallel}$ to ${A}^{\parallel} \sim {\delta E^{k \perp}}^{1/2} \sim \Delta \varphi^{1/2}$ as observed in (c).
Therefore, in the stressed system, we identify $\omega^\ast$ as the frequency-point where $\delta E^{k \parallel}$ becomes comparable to $\delta E^{k \perp}$.
Even though the transverse energy, $\delta E^{k \perp} \sim \Delta \varphi$, becomes very small close to the transition point ($\Delta \varphi \ll 1$), it cannot be neglected in the low $\omega$ regime below $\omega^\ast$, $\omega < \omega^\ast$.

{\bf Net displacement $\boldsymbol{{e}^{k \parallel}_\text{net}}$.}
The net displacement ${e}^{k \parallel}_\text{net}$ in Fig.~\ref{fig.vibration}(e), which is roughly two orders of magnitude smaller than the absolute displacement ${e}^{k \parallel}$, shows a similar $\omega$-dependence as ${e}^{k \parallel}$.
In particular, ${e}^{k \parallel}_\text{net}$ similarly exhibits a functional crossover at $\omega^\ast$, from ${e}^{k \parallel}_\text{net} \sim \omega$ to $\sim \omega^0$.
As $\omega \rightarrow 0$, ${e}^{k \parallel}_\text{net} \rightarrow A^{\parallel}_\text{net}$, which depends on $\varphi$ as $A^{\parallel}_\text{net} \sim \Delta \varphi^{1/2}$ in the same manner as $A^{\parallel}$.
Thus, we conclude that as for ${e}^{k \parallel}$, the crossover in ${e}^{k \parallel}_\text{net}$ at $\omega=\omega^\ast$ is also controlled by the competition between the two mode energies, $\delta E^{k \parallel}$ and $\delta E^{k \perp}$.
However, we see a difference between ${e}^{k \parallel}$ and ${e}^{k \parallel}_\text{net}$ at high frequencies $\omega>\omega^h$: ${e}^{k \parallel}_\text{net}$ shows a crossover from ${e}^{k \parallel}_\text{net} \sim \omega$ to $\sim \omega^0$, while ${e}^{k \parallel}$ retains the scaling ${e}^{k \parallel} \sim \omega$ with no crossover.

In order to characterize the crossover in ${e}^{k \parallel}_\text{net}$ at $\omega=\omega^h$, we divide ${e}^{k \parallel}_\text{net}$ into two terms, ${e}^{k \parallel}_\text{com}$ and ${e}^{k \parallel}_\text{str}$, which originate from the compressing ($\vec{e}^{k \parallel}_{ij} \cdot \vec{n}_{ij}<0$) and the stretching ($\vec{e}^{k \parallel}_{ij} \cdot \vec{n}_{ij}>0$) motions, respectively;
\begin{equation}
\begin{aligned} \label{displace5}
{e}^{k \parallel}_\text{net} &= \left| \frac{1}{N^\text{ct}} \left( \sum_{\vec{e}^{k \parallel}_{ij}\cdot \vec{n}_{ij}<0} + \sum_{\vec{e}^{k \parallel}_{ij}\cdot \vec{n}_{ij}>0} \right) \vec{e}^{k \parallel}_{ij}\cdot \vec{n}_{ij} \right|,\\
& := \left| -{e}^{k \parallel}_\text{com} + {e}^{k \parallel}_\text{str} \right|,
\end{aligned}
\end{equation}
where ${e}^{k \parallel}_\text{com} > 0$ and ${e}^{k \parallel}_\text{str} > 0$ are both positive quantities.
The absolute displacement ${e}^{k \parallel}$ can be approximated by a sum of those two terms; ${e}^{k \parallel} \approx {e}^{k \parallel}_\text{com} + {e}^{k \parallel}_\text{str}$.
We have confirmed that below $\omega^h$, the two terms increase with $\omega$, with different rates, i.e., ${e}^{k \parallel}_\text{com} \approx R_\text{com} \omega$ and ${e}^{k \parallel}_\text{str} \approx R_\text{str} \omega$ ($R_\text{com} \neq R_\text{str}$), and as a result, the net value ${e}^{k \parallel}_\text{net}$ increases as ${e}^{k \parallel}_\text{net} \approx \left| R_\text{str}-R_\text{com} \right| \omega$.
On the other hand, above $\omega^h$, they increase at the same rate, $R_\text{com} \approx R_\text{str} \approx R$, so that the net value does not vary with $\omega$.
The absolute ${e}^{k \parallel}$ increases as ${e}^{k \parallel} \approx (R_\text{str} + R_\text{com}) \omega$, both below and above $\omega^h$.
Therefore, we conclude that the crossover in ${e}^{k \parallel}_\text{net}$ at $\omega = \omega^h$ is determined by the balance between the compressing (${e}^{k \parallel}_\text{com}$) and the stretching (${e}^{k \parallel}_\text{str}$) motions.
The net displacement ${e}^{k \parallel}_\text{net}$ exhibits two crossovers at $\omega^\ast$ and $\omega^h$, such that the three regimes defined in $g(\omega)$~\cite{Silbert_2009,Xu_2010,Silbert_2005} can be distinguished by the scaling-behaviors of ${e}^{k \parallel}_\text{net}$ as
\begin{equation}
\begin{aligned} \label{resultvs2}
& {e}^{k \parallel}_\text{net} \sim
\left\{ \begin{aligned}
& \omega^0 & (\omega > \omega^h), \\
& \omega   & (\omega^\ast < \omega < \omega^h), \\
& \omega^0 & (\omega < \omega^\ast).
\end{aligned} \right. \\
\end{aligned}
\end{equation}

{\bf Comparison to unstressed system.}
Finally we look at the unstressed system in right panels of Fig.~\ref{fig.vibration}.
Above $\omega^\ast$, where $\delta E^{k \parallel}$ controls the total mode energy $\delta E^k$ in the stressed system, the unstressed system exhibits the same behaviors and power-law scalings as the stressed system.
However, since $\delta E^{k \perp} \equiv 0$ and $\delta E^{k \parallel} \equiv \delta E^k$, the unstressed system shows no crossover at $\omega = \omega^\ast$, and no distinct behaviors between $\omega>\omega^\ast$ and $\omega<\omega^\ast$.
Therefore, although the unstressed system is a good approximation to the original stressed system, the low $\omega < \omega^\ast$ modes (low energy modes) behave differently between the two systems.

\begin{figure}[t]
\centering
\includegraphics[width=0.475\textwidth]{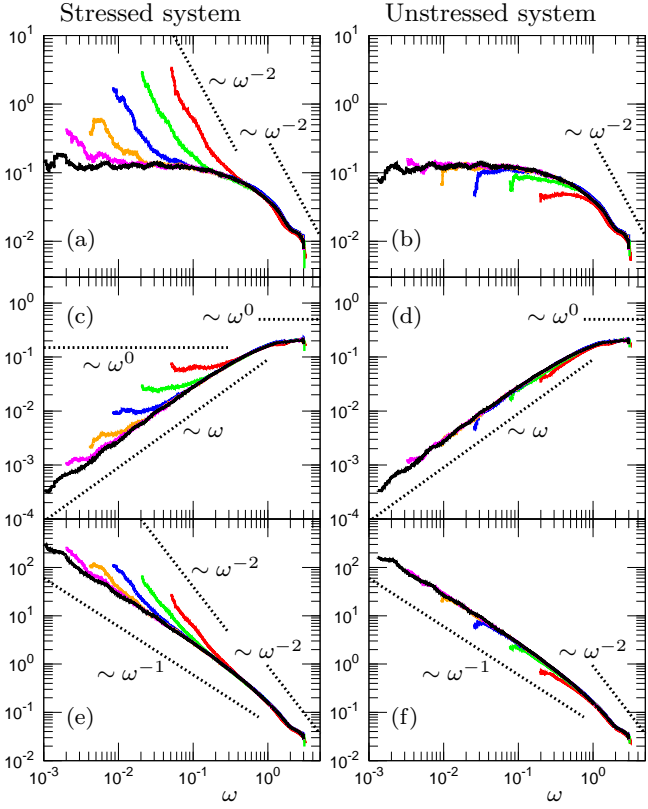}
\vspace*{0mm}
\caption{\label{fig.decomposition}
(Color online) Eigenmode decomposition of non-affine \textit{bulk} modulus in the stressed (left panels) and the unstressed (right panels) systems.
We plot the non-affine modulus $K_N^k$ in (a),(b), force field $\left| \Sigma_K^k \right|$ in (c),(d), and the non-affine displacement field $\left| \delta R^k_{\text{na} K} \right|$ in (e),(f), as functions of the eigenfrequency $\omega$.
The values are averaged over the frequency bins of $\log_{10} \omega^k \in [\log_{10} \omega-\Delta \omega/2,\log_{10} \omega + \Delta \omega/2]$ with $\Delta \omega = 0.07$.
The different lines indicate different packing fractions, $\Delta \varphi = 10^{-1}$ (red), $10^{-2}$ (green), $10^{-3}$ (blue), $10^{-4}$ (orange), $10^{-5}$ (magenta), $10^{-6}$ (black), from right to left or from top to bottom.
The detailed description of presented quantities is given in Sec.~\ref{sec.moduli}.}
\end{figure}

\begin{figure}[t]
\centering
\includegraphics[width=0.475\textwidth]{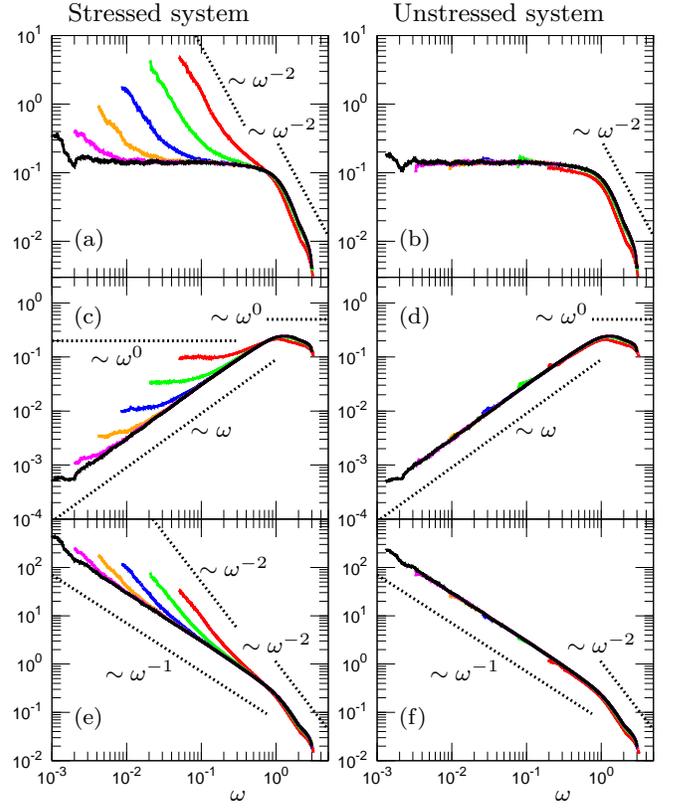}
\vspace*{0mm}
\caption{\label{fig.decompositiona}
(Color online) Eigenmode decomposition of non-affine \textit{shear} modulus in the stressed (left panels) and the unstressed (right panels) systems.
We plot the non-affine modulus $G_N^k$ in (a),(b), force field $\left| \Sigma_G^k \right|$ in (c),(d), and the non-affine displacement field $\left| \delta R^k_{\text{na} G} \right|$ in (e),(f), as functions of the eigenfrequency $\omega$.
See the caption of Fig.~\ref{fig.decomposition}.}
\end{figure}

\begin{figure}[t]
\centering
\includegraphics[width=0.48\textwidth]{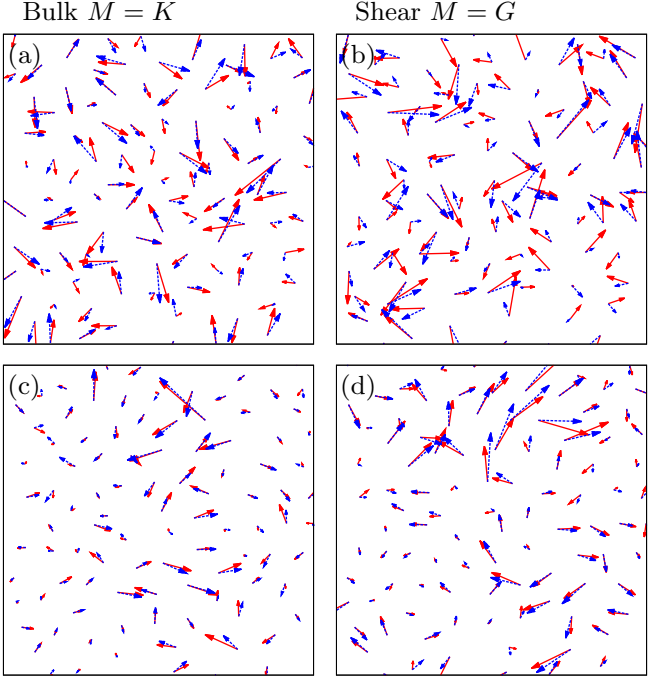}
\vspace*{0mm}
\caption{\label{fig.picture}
(Color online) Spatial maps of the force field $\vec{\Sigma}_M$ ((a),(b)) and the non-affine displacement field $\delta \vec{R}_{\text{na} M}$ ((c),(d)) in real space, corresponding to bulk $M=K$ (left panels) and shear $M=G$ (right panels) deformations.
The packing fraction is $\Delta \varphi = 10^{-5}$.
We plot the vector fields at a fixed plane within the packing of thickness $\approx 1 [\sigma]$, which includes around $100$ particles ($10\%$ of all the particles).
$\vec{\Sigma}_M$ and $\delta \vec{R}_{\text{na} M}$ are formulated as a superposition of the eigenvectors $\vec{e}^k$ weighted by the components of $\Sigma_M^k$ and $\delta {R}_{\text{na} M}^k$, respectively (see Eqs.~(\ref{modulus7}) and~(\ref{modulus7a})).
In the figure, we show the fields obtained by a summation of all the eigenmodes $k=1,2,...,3N-3$ (red solid vectors), and those obtained by a partial summation over $\omega^k > \omega^h$ for $\vec{\Sigma}_M$, and $\omega^k < \omega^\ast$ for $\delta \vec{R}_{\text{na} M}$ (blue dashed vectors).}
\end{figure}

\begin{figure}[t]
\centering
\includegraphics[width=0.48\textwidth]{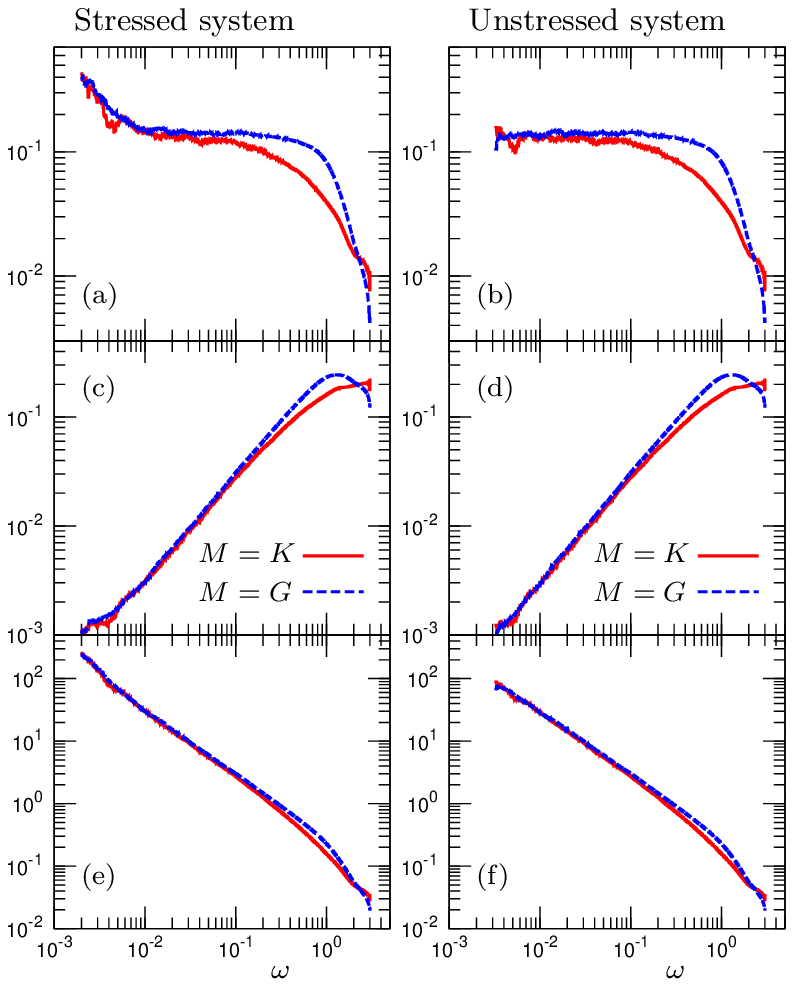}
\vspace*{0mm}
\caption{\label{fig.decomposition2}
(Color online) Comparison of $M_N^k, \left| \Sigma_M^k \right|, \left| \delta R^k_{\text{na} M} \right|$ between the bulk $M=K$ (red solid line) and the shear $M=G$ (blue dashed line) moduli, for the stressed (left panels) and the unstressed (right panels) systems.
We plot $M_N^k$ in (a),(b), $\left| \Sigma_M^k \right|$ in (c),(d), and $\left| \delta R^k_{\text{na} M} \right|$ in (e),(f), as functions of the eigenfrequency $\omega$.
The packing fraction is $\Delta \varphi = 10^{-5}$.
The data are same as those presented in Figs.~\ref{fig.decomposition} and~\ref{fig.decompositiona}.}
\end{figure}

\subsection{Eigenmode decomposition of non-affine moduli} \label{sec.nonaffine1}
In this section, we study the non-affine modulus $M_N$, which is decomposed by eigenmode $k$ contribution, $M_{N}^{k}$ ($k=1,2,...,3N-3$), as in Eq.~(\ref{modulus10}).
Each component $M_{N}^{k}$ is formulated as the product of force $\Sigma_M^k$ and non-affine displacement $\delta {R}_{\text{na} M}^k$, and thus can be interpreted as an energy relaxation by the eigenmode $k$ excitation during non-affine deformation process.
The values of $M_{N}^{k}$, $\left| \Sigma_M^k \right|$, $\left| \delta {R}_{\text{na} M}^k \right|$ are presented as functions of the eigenfrequency $\omega$, for the range of packing fraction, $\Delta \varphi = 10^{-1}$ to $10^{-6}$, in Fig.~\ref{fig.decomposition} for the bulk $M=K$ and Fig.~\ref{fig.decompositiona} for the shear $M=G$.
Note that since $M_{N}^{k}$ is positive for all the modes $k$, $M_{N}^{k} = \left| \Sigma_M^k \right| \times \left| \delta {R}_{\text{na} M}^k \right|$ holds.
The presented values are averaged over the frequency bins of $\log_{10} \omega^k \in [\log_{10} \omega-\Delta \omega/2,\log_{10} \omega + \Delta \omega/2]$ with $\Delta \omega = 0.07$.

{\bf Eigenmode contribution $\boldsymbol{M_{N}^{k}}$.}
We first focus on the stressed system in the left panels of Figs.~\ref{fig.decomposition} and~\ref{fig.decompositiona}.
Like the vibrational modes in Fig.~\ref{fig.vibration}, the non-affine modulus $M_N^k$, in (a), also shows three distinct frequency regimes; (i) intermediate $\omega^\ast_M < \omega < \omega^h_M$ regime, (ii) low $\omega < \omega^\ast_M$ regime, and (iii) high $\omega > \omega^h_M$ regime.
At intermediate frequencies, $\omega^\ast_M < \omega < \omega^h_M$, $M_N^k$ is practically $\omega$-independent and shows a plateau.
In the low-frequency regime, $\omega < \omega^\ast_M$, $M_N^k$ increases from the plateau value as $M_N^k \sim \omega^{-2}$.
Finally, in the high-frequency regime, $\omega > \omega^h_M$, $M_N^k$ drops and decreases as $\omega \rightarrow \omega_\text{max} \simeq 3$.
Here, we remark that the bulk modulus $K_N^k$ is not strictly a plateau in the intermediate regime but slightly decreases at higher $\omega$, so that we cannot cleanly identify $\omega_K^\ast$ at higher $\Delta \varphi$, and $\omega_K^h$.
Thus, we determined $\omega^\ast_{K}$ only for the lower $\Delta \varphi \le 5\times 10^{-3}$, and did not identify a specific $\omega_{K}^h$.
Whereas the shear modulus $G_N^k$ shows a clear plateau region, and we can determine both $\omega^\ast_G$ and $\omega^h_G$ without ambiguity.
We discuss this difference between $K_N^k$ and $G_N^k$ at the end of this section, but here we emphasize that at a qualitative level, $K_N^k$ can also be divided into three regimes as described above.
In order to check if the crossover points coincide between the vDOS $g(\omega)$ and $M_N^k$, we compare $\omega^\ast, \omega^h$ from $g(\omega)$, to $\omega^\ast_{M}, \omega^h_{M}$ from $M_N^k$ in Fig.~\ref{fig.frequency}(b).
Figure~\ref{fig.frequency}(b) indeed demonstrates that $g(\omega)$ and $M_N^k$ indicate the same crossover frequencies: $\omega^\ast \simeq \omega^\ast_M \sim \Delta \varphi^{1/2}$ and $\omega^h \simeq \omega^h_M \simeq 1.0$.

{\bf Force $\boldsymbol{\left| \Sigma_M^k \right|}$ and non-affine displacement $\boldsymbol{\left| \delta {R}_{\text{na} M}^k \right|}$.}
We turn to the force $\left| \Sigma_M^k \right|$ in (c) of Figs.~\ref{fig.decomposition} and~\ref{fig.decompositiona}, and the non-affine displacement $\left| \delta {R}_{\text{na} M}^k \right|$ in (e).
As in Eq.~(\ref{displace4}), $\left| \Sigma_M^k \right|$ and $\left| \delta {R}_{\text{na} M}^k \right|$ are directly related to the net (compressing/stretching) displacement ${e}^{k \parallel}_\text{net}$.
Indeed, we observe the following power-law behaviors of $\left| \Sigma_M^k \right|, \left| \delta {R}_{\text{na} M}^k \right|, M_N^k = \left| \Sigma_M^k \right| \times \left| \delta {R}_{\text{na} M}^k \right|$;
\begin{equation}
\begin{aligned} \label{resultenedis}
& \left| \Sigma_M^k \right| \sim {e}^{k \parallel}_\text{net} \sim
\left\{ \begin{aligned}
& \omega^0   & (\omega > \omega^h), \\
& \omega   & (\omega^\ast < \omega < \omega^h), \\
& \omega^0 & (\omega < \omega^\ast),
\end{aligned} \right. \\
& \left| \delta {R}_{\text{na} M}^k \right| \sim \frac{ {e}^{k \parallel}_\text{net} }{ {\omega}^{2} } \sim
\left\{ \begin{aligned}
& \omega^{-2}   & (\omega > \omega^h), \\
& \omega^{-1} & (\omega^\ast < \omega < \omega^h), \\
& \omega^{-2} & (\omega < \omega^\ast),
\end{aligned} \right. \\
& M_N^k \sim \frac{ { {e}^{k \parallel}_\text{net} }^2 }{ {\omega}^{2} } \sim
\left\{ \begin{aligned}
& \omega^{-2}   & (\omega > \omega^h), \\
& \omega^{0}  & (\omega^\ast < \omega < \omega^h), \\
& \omega^{-2} & (\omega < \omega^\ast),
\end{aligned} \right.
\end{aligned}
\end{equation}
all of which are consistent with the behavior of ${e}^{k \parallel}_\text{net}$ in Eq.~(\ref{resultvs2}).
As $\omega \rightarrow 0$, ${e}^{k \parallel}_\text{net} \rightarrow A^{\parallel}_\text{net} \sim \Delta \varphi^{1/2}$, leading to $\left| \Sigma_M^k \right| \sim A^{\parallel}_\text{net} \sim \Delta \varphi^{1/2}$, $\left| \delta {R}_{\text{na} M}^k \right| \sim A^{\parallel}_\text{net} \omega^{-2} \sim \Delta \varphi^{1/2} \omega^{-2}$, and $M_N^k \sim {A^{\parallel 2}_\text{net}} \omega^{-2} \sim \Delta \varphi \omega^{-2}$.
Therefore, all of $\left| \Sigma_M^k \right|, \left| \delta {R}_{\text{na} M}^k \right|, M_N^k$ follow the net displacement ${e}^{k \parallel}_\text{net}$.
Particularly, their crossovers at $\omega^\ast$ are controlled by the competition between the compressing/stretching $\delta E^{k \parallel}$ and sliding $\delta E^{k \perp}$ energies, whereas those at $\omega^h$ are determined by the balance between the compressing ${e}^{k \parallel}_\text{com}$ and stretching ${e}^{k \parallel}_\text{str}$ motions.

{\bf Comparison to unstressed system.}
When comparing the stressed system (left panels of Figs.~\ref{fig.decomposition} and~\ref{fig.decompositiona}) to the unstressed system (right panels), both systems show the same behaviors of $\left| \Sigma_M^k \right|,\left| \delta {R}_{\text{na} M}^k \right|,M_N^k$, at $\omega > \omega^\ast$, particularly the same power-law scalings.
However, since the unstressed system shows no crossover in ${e}^{k \parallel}_\text{net}$ (and ${e}^{k \parallel}$, $\delta E^{k \parallel}$) at $\omega=\omega^\ast$, as discussed in the previous Sec.~\ref{sec.vibrationalstate}, it retains the same behaviors of $\left| \Sigma_M^k \right|,\left| \delta {R}_{\text{na} M}^k \right|,M_N^k$ at $\omega^\ast < \omega < \omega^h$ down to $\omega = 0$, i.e., at $0 <\omega <\omega^h$.
Thence, below $\omega^\ast$, the two systems show distinct behaviors and scalings in their vibrational modes as well as the non-affine elastic moduli.
This result is a direct consequence that the transverse energy $\delta E^{k \perp}$ in the stressed system is effective below $\omega^\ast$, but negligible above $\omega^\ast$.

{\bf Physical interpretation of $\boldsymbol{M_{N}^{k}}$.}
We can interpret our results of $M_N^k = \left| \Sigma_M^k \right| \times \left| \delta {R}_{\text{na} M}^k \right|$ in Figs.~\ref{fig.decomposition} and~\ref{fig.decompositiona}, and Eq.~(\ref{resultenedis}), in terms of energy relaxation during the non-affine deformation process.
At the highest frequencies, $\omega > \omega^h$, there exists a bunch of closely spaced, localized eigenmodes of a sufficiently high energy that they are only weakly activated.
As a result, their associated non-affine displacement fields are small, leading to minimal energy relaxation and $M_N^k$.
At intermediate frequencies, $\omega^\ast < \omega < \omega^h$, the modes are of lower energies and are more readily excited.
As a result, the nonaffine displacement grows as $\left| \delta {R}_{\text{na} M}^k \right| \sim \omega^{-1}$, whereas at the same time, the force $\left| \Sigma_M^k \right| \sim \omega$ becomes smaller with decreasing frequency.
These two competing effects balance, resulting in the constant, plateau value of energy relaxation, $M^k_N \sim \omega^0$.
Finally, at the low end of the frequency spectrum, $\omega < \omega^\ast$, for the stressed system, the stress, $\sim \phi'(r_{ij})$, enhances the force $\left| \Sigma_M^k \right|$ and drives the non-affine displacement $\left| \delta {R}_{\text{na} M}^k \right|$.
Since the stress term, $\sim \phi'(r_{ij})$, reduces the mode energy by $\delta E^{k \perp}$ (see Eq.~(\ref{vs1})), the compressing/stretching energy $\delta E^{k \parallel}$ compensates this destabilization of the system, leading to the larger value of ${e}^{k \parallel}_\text{net}$ (and also ${e}^{k \parallel}$) and then the enhancements of $\left| \Sigma_M^k \right|$ and $\left| \delta {R}_{\text{na} M}^k \right|$.
As a result, the energy relaxation grows with decreasing $\omega$ as $M^k_N \sim \omega^{-2}$.
While, the unstressed system with zero stress, $\sim \phi'(r_{ij}) \equiv 0$, has a constant energy relaxation, $M_N^k \sim \omega^0$, even at $\omega < \omega^\ast$, as it does at $\omega^\ast < \omega < \omega^h$.

{\bf Spatial structures of $\boldsymbol{\left| \Sigma_M^k \right|}$ and $\boldsymbol{\left| \delta {R}_{\text{na} M}^k \right|}$.}
As reported by Maloney and Lema\^{i}tre~\cite{Maloney_2004,Maloney2_2006,Maloney_2006,Lemaitre_2006}, the force field $\vec{\Sigma}_M$ exhibits a random structure (without any apparent spatial correlation) in real space, while the non-affine displacement field $\delta \vec{R}_{\text{na} M}$ shows a vortex-like structure (with apparent long-range spatial correlation).
Indeed, such features are observed in Fig.~\ref{fig.picture}, where $\vec{\Sigma}_M$ and $\delta \vec{R}_{\text{na} M}$ are visualized in real space, at a fixed plane within a slice of thickness of a particle diameter.
As in Eqs.~(\ref{modulus7}) and~(\ref{modulus7a}), the real-space structures of $\vec{\Sigma}_M$ and $\delta \vec{R}_{\text{na} M}$ are constructed as a superposition of the eigenvectors $\vec{e}^k$ weighted by the components of $\Sigma_M^k$ and $\delta {R}_{\text{na} M}^k$.
Figure~\ref{fig.picture} also compares the total contributions (red solid vectors) to those obtained by a partial summation over $\omega^k > \omega^h$ for $\vec{\Sigma}_M$, and $\omega^k < \omega^\ast$ for $\delta \vec{R}_{\text{na} M}$ (blue dashed vectors).
It is seen that the partial summations can well reproduce the true fields (full summations) of $\vec{\Sigma}_M$ and $\delta \vec{R}_{\text{na} M}$.
Therefore, our results indicate that the eigenvectors $\vec{e}^k$ at high frequencies $\omega^k > \omega^h$, which are highly localized fields \cite{Silbert_2009,Xu_2010}, mainly contribute to the random structure of $\vec{\Sigma}_M$ (Fig.~\ref{fig.picture}(a),(b)).
While the vortex-like, structure of $\delta \vec{R}_{\text{na} M}$ (Fig.~\ref{fig.picture}(c),(d)) comes from the transverse fields with vortex features apparent in the eigenvectors $\vec{e}^k$ at low frequencies $\omega^k < \omega^\ast$ \cite{Silbert_2005,Silbert_2009}.
Here we should remark that on approach to the transition point, $\Delta \varphi \rightarrow 0$ and $\omega^\ast \rightarrow 0$, the contributions to $\delta \vec{R}_{\text{na} M}$ at $\omega^k < \omega^\ast$ become less and less, and finally the modes at $\omega > \omega^\ast$ also start to play a role in determining $\delta \vec{R}_{\text{na} M}$.

{\bf Comparison between bulk $\boldsymbol{M=K}$ and shear $\boldsymbol{M=G}$ moduli.}
We close this section with a comparison of $M_{N}^{k},\left| \Sigma_M^k \right|,\left| \delta {R}_{\text{na} M}^k \right|$ between the bulk $M=K$ (Fig.~\ref{fig.decomposition}) and shear $M=G$ (Fig.~\ref{fig.decompositiona}) moduli.
All of $M_{N}^{k}$, $\left| \Sigma_M^k \right|$, $\left| \delta {R}_{\text{na} M}^k \right|$ show similar behaviors and power-law scalings between $M=K$ and $G$, for both the stressed and unstressed systems.
However, we observe some differences: At $\omega^\ast < \omega < \omega^h$, $G_N^k$ shows a clear plateau, while $K_N^k$ slightly depends on $\omega$.
We focus on these differences in Fig.~\ref{fig.decomposition2}, where we compare $M_N^k, \left| \Sigma_M^k \right|,\left| \delta {R}_{\text{na} M}^k \right|$ between $M=K$ and $G$.
At lower frequencies $\omega \lesssim 10^{-1}$, the quantities coincide well between $M=K$ and $G$
\footnote{Here we remark that the average values of $\left| \Sigma_M^k \right|,\left| \delta {R}_{\text{na} M}^k \right|,M_N^k$ over different realizations and frequency shells coincide between $M=K$ and $G$ at $\omega \lesssim 10^{-1}$.
However, those quantities of one realization and one mode $k$ show different values between $M=K$ and $G$.
Thus, the vector fields of $\vec{\Sigma}_M$ and $\delta \vec{R}_{\text{na} M}$ of one realization are different between $M=K$ and $G$, even if they are constructed by a partial summation over $\omega^k \lesssim 10^{-1}$.
This point is indeed seen by comparing $\delta \vec{R}_{\text{na} K}$ and $\delta \vec{R}_{\text{na} G}$ in Fig.~\ref{fig.picture}(c),(d), where we plot $\delta \vec{R}_{\text{na} M}$ constructed by the modes with $\omega < \omega^\ast (\ll 10^{-1})$ (see the blue dashed vectors).},
whereas at higher frequencies $\omega \gtrsim 10^{-1}$, they are larger for $G$ than for $K$.
Here we note that $K_N^k$ starts to deviate from its plateau value at $\omega \approx 10^{-1}$.
Thus, eigenmodes with $\omega \gtrsim 10^{-1}$ are excited more under shear deformation than under compressional deformation, which results in more energy relaxation and a larger non-affine modulus $G_N$ than $K_N$.
As we will see in Eq.~(\ref{nonaffine7}) in the next section, the critical value of $G_{Nc} \simeq 0.24$ is larger than $K_{Nc} \simeq 0.15$, which comes from the eigenmodes contributions at $\omega \gtrsim 10^{-1}$.

Ellenbroek \textit{et. al.}~\cite{Ellenbroek_2006,Ellenbroek_2009,Ellenbroek2_2009} have demonstrated a distinction in non-affine responses under compression and shear: The non-affine response under shear is considered to be governed by more floppy-like motions than that under compression.
From their result, we might expect that the floppy-like, vibrational modes at low frequencies are more enhanced under shear than under compression.
However, our results indicate that this issue is more subtle and involves an interplay between the modes over the entire vibrational spectrum.
While it is true that the large-scale nonaffine field, $\delta \vec{R}_{\text{na} M}$, comes from the lower frequency portion of the spectrum for both compression and shear, the difference between them appears at relatively high frequencies $\omega \gtrsim 10^{-1}$, \textit{not} really low frequencies (for the example, $\Delta\varphi = 10^{-5}$, shown in Fig.~\ref{fig.decomposition2}).
Therefore, if one associates ``floppiness'' with more non-affine or softer under shear than under compression, this is not a property restricted to just the low frequency modes.

\begin{figure}[t]
\centering
\includegraphics[width=0.48\textwidth]{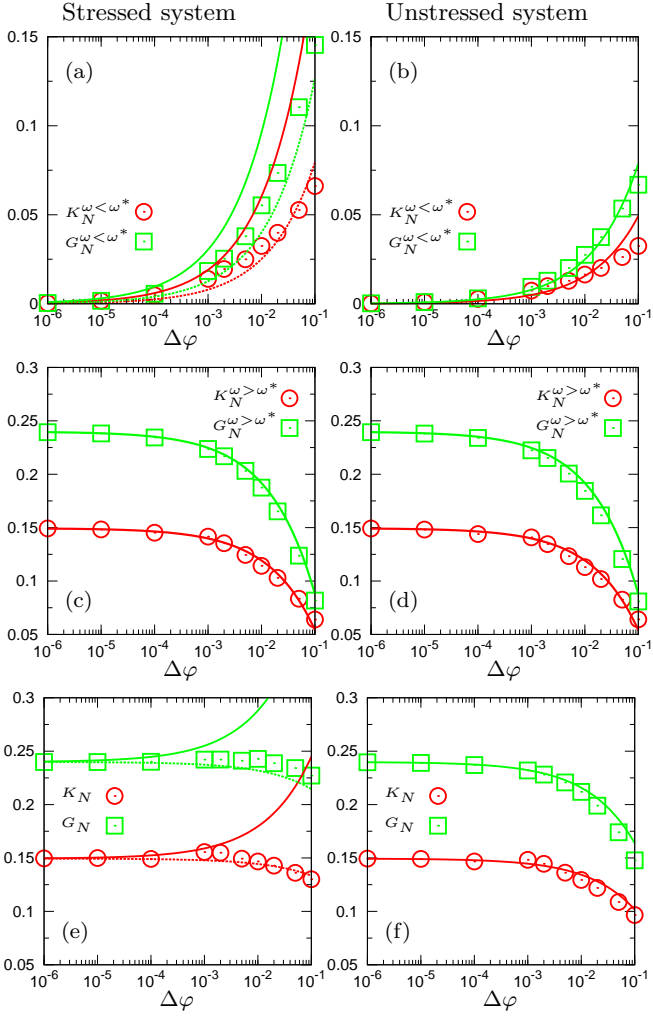}
\vspace*{0mm}
\caption{\label{fig.nonaffine}
(Color online)
Non-affine moduli $K_N, G_N$ in the stressed (left panels) and the unstressed (right panels) systems.
We plot $K_{N}^{\omega < \omega^\ast}, G_{N}^{\omega < \omega^\ast}$ in (a),(b), $K_{N}^{\omega > \omega^\ast}, G_{N}^{\omega > \omega^\ast}$ in (c),(d), and total $K_{N}, G_{N}$ in (e),(f).
In the figures, we compare numerical values presented in Fig.~\ref{fig.phidependence}(a),(b) (symbols), to the formulations (solid lines) which are described in the main text (see Eqs.~(\ref{nonaffine2}), (\ref{nonaffine6}), (\ref{nonaffine9})).
Note that the numerical values of $M_{N}^{\omega < \omega^\ast}$ and $M_{N}^{\omega > \omega^\ast}$ are obtained by replacing $\sum_{k=1}^{3N-3}$ in Eq.~(\ref{modulus10}) with partial summations, $\sum_{\omega^k < \omega^\ast}$ and $\sum_{\omega^k > \omega^\ast}$, respectively.
In (a) and (e) for the stressed system, dashed lines indicate the formulation where we use the exponents of $a=1.5$ and $b=1.3$ (see Eqs.~(\ref{nonaffine2x}), (\ref{nonaffine9x})).}
\end{figure}

\subsection{Formulation of non-affine moduli} \label{sec.nonaffine2}
Based on observations in the previous Secs.~\ref{sec.vibrationalstate} and~\ref{sec.nonaffine1}, we attempt to formulate the non-affine modulus $M_N=K_{N},G_N$.
Following Refs.~\cite{Lemaitre_2006,Zaccone_2011}, we assume that $M_{N}^{k}$ (also $\left| \Sigma_M^k \right|, \left| \delta {R}_{\text{na} M}^k \right|$) is a self-averaged quantity: In the thermodynamics limit $N \rightarrow \infty$, $M_{N}^{k}$ converges to a well-defined continuous function of $\omega$, i.e., $M_{N}^{k}(\omega)$, which can be then obtained by averaging over the frequency shells and different realizations, as we have done in Figs.~\ref{fig.decomposition} and~\ref{fig.decompositiona} for $M_{N}^{k} = K_{N}^{k}$ and $G_{N}^{k}$, respectively.
Thus we replace the summation, $\sum_{k=1}^{3N-3}$, in $M_N$ of Eq.~(\ref{modulus10}) by the integral operator, $\int d\omega (3N-3) g(\omega) \simeq \int d\omega 3N g(\omega)$;
\begin{equation} \label{nonaffine1x}
M_{N} = \frac{1}{V} \sum_{k=1}^{3N-3} M_{N}^{k} = 3\hat{\rho} \int d\omega g(\omega) M_{N}^k(\omega),
\end{equation}
where we note $(3N-3) g(\omega) \simeq 3N g(\omega)$ is the total number of the eigenmodes $k$ per unit frequency at $\omega$.
We then separate $M_N$ into two terms, by dividing the integral regime into $\omega < \omega^\ast$ and $\omega > \omega^\ast$;
\begin{equation}
\begin{aligned} \label{nonaffine1}
M_{N} &= 3\hat{\rho} \left( \int_{\omega<\omega^\ast} d\omega + \int_{\omega>\omega^\ast} d\omega \right) g(\omega) M_{N}^k(\omega),\\
&:= M_{N}^{\omega < \omega^\ast} + M_{N}^{\omega > \omega^\ast}.
\end{aligned}
\end{equation}
In the following, we deal with those two terms in turn.

{\bf Formulation of $\boldsymbol{M_{N}^{\omega < \omega^\ast}}$.}
For $\omega < \omega^\ast$, we suppose a Debye-like density of states, as observed in Fig.~\ref{fig.vibration}(a),(b);
\begin{equation} \label{assume1}
g(\omega) = g^\ast \left( \frac{\omega}{\omega^\ast} \right)^a,
\end{equation}
where $g^\ast$ is the plateau value of $g(\omega)$, and the exponent $a$ depends on the stressed or unstressed systems;
\begin{equation}
a = \left\{ \begin{aligned}
& \frac{3}{2} & \text{(stressed)}, \\
& 1           & \text{(unstressed)}.
\end{aligned} \right.
\end{equation}
In addition, from Figs.~\ref{fig.decomposition}(a),(b) and~\ref{fig.decompositiona}(a),(b), we also reasonably assume
\begin{equation} \label{assume2}
M_{N}^k(\omega) = M_{N}^{\ast} \left( \frac{\omega}{\omega^\ast} \right)^{-b},
\end{equation}
where $M_{N}^{\ast}$ represents the plateau value of $M_{N}^k (\omega)$, and the exponent $b$ is
\begin{equation}
b = \left\{ \begin{aligned}
& 2 & \text{(stressed)}, \\
& 0 & \text{(unstressed)}.
\end{aligned} \right.
\end{equation}
On performing the integral $\int_{\omega < \omega^\ast} d\omega$ in Eq.~(\ref{nonaffine1}), we obtain $M_{N}^{\omega < \omega^\ast}$ as
\begin{equation}
\begin{aligned} \label{nonaffine2}
M_{N}^{\omega < \omega^\ast} &= \left( \frac{ 1 }{a-b+1} \right) 3\hat{\rho} g^\ast M_N^{\ast} {\omega^\ast}, \\
& = \left\{ \begin{aligned}
& 6 \hat{\rho}_c g^\ast M_N^{\ast} {\omega^\ast} + \mathcal{O}(\Delta \varphi) & \text{(stressed)}, \\
& \frac{3}{2} \hat{\rho}_c g^\ast M_N^{\ast} {\omega^\ast} + \mathcal{O}(\Delta \varphi) & \text{(unstressed)}.
\end{aligned} \right.
\end{aligned}
\end{equation}
Note that in the stressed case, the integrand function, $g(\omega) M_{N}^k(\omega) \sim \omega^{a-b} \sim \omega^{-1/2}$, diverges to $+\infty$ as $\omega \rightarrow 0$, but its integral over $\omega=0$ to $\omega^\ast$ converges to a finite value.
As $\Delta \varphi \rightarrow 0$, $\omega^\ast$ goes to zero, i.e., the Debye-like region disappears, and $M_{N}^{\omega < \omega^\ast}$ vanishes as $M_{N}^{\omega < \omega^\ast} \sim \omega^\ast \sim \Delta \varphi^{1/2} \rightarrow 0$.

{\bf Formulation of $\boldsymbol{M_{N}^{\omega > \omega^\ast}}$.}
Next we consider the integral $\int_{\omega > \omega^\ast} d\omega$ in Eq.~(\ref{nonaffine1}), i.e., $M_{N}^{\omega > \omega^\ast}$.
Since $g(\omega)$ and $M_{N}^k(\omega)$ are independent of $\Delta \varphi$ at $\omega > \omega^h$, the integral of $\int_{\omega > \omega^h} d\omega$ gives a constant value as;
\begin{equation} \label{nonaffine4}
\int_{\omega > \omega^h} d\omega g(\omega) M_{N}^k(\omega) = M_{N}^h \ \text{(constant)}.
\end{equation}
In the regime of $\omega^\ast < \omega < \omega^h$, both $g(\omega)$ and $M_{N}^k(\omega)$ show the plateau, thus we formulate
\begin{equation} \label{nonaffine5}
\int_{\omega^\ast < \omega <\omega^h} d\omega g(\omega) M_{N}^k(\omega) = g^\ast M_{N}^{\ast} \left( \omega^h-\omega^\ast \right).
\end{equation}
Therefore, we arrive at
\begin{equation} \label{nonaffine6}
\begin{aligned}
M_{N}^{\omega > \omega^\ast} &= 3\hat{\rho} \left(M_{N}^h + g^\ast M_{N}^{\ast} \omega^h \right) - 3\hat{\rho} g^\ast M_{N}^{\ast} \omega^\ast, \\
 &= M_{Nc} - 3\hat{\rho}_c g^\ast M_{N}^{\ast} \omega^\ast + \mathcal{O}(\Delta \varphi),
\end{aligned}
\end{equation}
where $M_{Nc} = 3\hat{\rho}_c \left(M_{N}^h + g^\ast M_{N}^{\ast} \omega^h \right)$ is the critical value at $\varphi_c$.
Thus, as $\Delta \varphi \rightarrow 0$ and $\omega^\ast \rightarrow 0$, the plateau region extends down to zero frequency, and $M_{N}^{\omega > \omega^\ast} \rightarrow M_{Nc}$.
We note that $M_{Nc}$ is the critical value not only for $M_{N}^{\omega > \omega^\ast}$ but also for the total non-affine modulus $M_N$, since $M_{N}^{\omega < \omega^\ast} \rightarrow 0$ as $\Delta \varphi \rightarrow 0$.

{\bf Summation of $\boldsymbol{M_{N}^{\omega < \omega^\ast}}$ and $\boldsymbol{M_{N}^{\omega > \omega^\ast}}$.}
Finally we sum up two terms of $M_{N}^{\omega < \omega^\ast}$ and $M_{N}^{\omega > \omega^\ast}$, and obtain the total modulus $M_N$ as
\begin{equation}
\begin{aligned} \label{nonaffine9}
M_{N} &= M_{Nc} - \left( \frac{ a-b }{a-b+1} \right) 3\hat{\rho}_c g^\ast M_{N}^{\ast} \omega^\ast + \mathcal{O}(\Delta \varphi), \\
& = \left\{ \begin{aligned}
& M_{Nc} + 3 \hat{\rho}_c g^\ast M_N^{\ast} {\omega^\ast} + \mathcal{O}(\Delta \varphi) & \text{(stressed)}, \\
& M_{Nc} - \frac{3}{2} \hat{\rho}_c g^\ast M_N^{\ast} {\omega^\ast} + \mathcal{O}(\Delta \varphi) & \text{(unstressed)}.
\end{aligned} \right.
\end{aligned}
\end{equation}
Here we note that $\omega^\ast \sim \Delta \varphi^{1/2}$ is the leading order term of $M_{N}^{\omega < \omega^\ast}$, $M_{N}^{\omega > \omega^\ast}$, $M_{N}$ in Eqs.~(\ref{nonaffine2}), (\ref{nonaffine6}), (\ref{nonaffine9}), respectively.
We have extracted the values of parameters in Eq.~(\ref{nonaffine9}), from data presented in Figs.~\ref{fig.vibration}, \ref{fig.decomposition}, and~\ref{fig.decompositiona};
\begin{equation}
\begin{aligned}
& g^\ast = 0.390,\qquad K_N^{\ast}=0.0740,\qquad G_N^{\ast}=0.118,\\
& K_N^h = 0.0135,\qquad G_N^h = 0.0219,
\end{aligned}
\end{equation}
which are common to the stressed and unstressed systems.
As mentioned in the previous Sec.~\ref{sec.nonaffine1} and Figs.~\ref{fig.decomposition} and~\ref{fig.decompositiona}, $G_N^k (\omega)$ shows a clear plateau over the intermediate frequency range, $\omega^\ast < \omega < \omega^h$, while $K_N^k (\omega)$ slightly depends on $\omega$.
Therefore, to take into account this dependence of $K_N^k (\omega)$, we determined the plateau value of $K_N^{\ast}$ as the average value of $K_N^{k} (\omega)$ over $\omega^\ast < \omega < \omega^h$ at the lowest packing fraction $\Delta \varphi = 10^{-6}$.
From the above values of parameters, we obtain the critical value, $M_{Nc}=3\hat{\rho}_c \left(M_{N}^h + g^\ast M_{N}^{\ast} \omega^h \right)$;
\begin{equation} \label{nonaffine7}
K_{Nc} \simeq 0.15, \qquad G_{Nc} \simeq 0.24.
\end{equation}

Figure \ref{fig.nonaffine} compares the simulation values (symbols) to the formulations of Eqs.~(\ref{nonaffine2}), (\ref{nonaffine6}), (\ref{nonaffine9}) (solid lines), for $M_{N}^{\omega < \omega^\ast}$ in (a),(b), $M_{N}^{\omega > \omega^\ast}$ in (c),(d), and the total $M_N$ in (e),(f).
We note that the simulation values of $M_{N}^{\omega < \omega^\ast}$ and $M_{N}^{\omega > \omega^\ast}$ are obtained by replacing $\sum_{k=1}^{3N-3}$ in Eq.~(\ref{modulus10}) with partial summations, $\sum_{\omega^k < \omega^\ast}$ and $\sum_{\omega^k > \omega^\ast}$, respectively.
It is seen that our formulation accurately captures $M_{N}^{\omega > \omega^\ast}$, while there is a discrepancy in $M_{N}^{\omega < \omega^\ast}$ of the stressed system (see Fig.~\ref{fig.nonaffine}(a)).
This discrepancy comes from the \textit{smooth} crossovers at $\omega = \omega^\ast$ in $g(\omega)$ and $M^k_N (\omega)$ (see Figs.~\ref{fig.vibration}(a), \ref{fig.decomposition}(a), \ref{fig.decompositiona}(a)), around which the assumptions of Eqs.~(\ref{assume1}) and~(\ref{assume2}) do not strictly hold.
In the unstressed system, there is a sharp crossover in $g(\omega)$ (Fig.~\ref{fig.vibration}(b)) and no crossover in $M^k_N (\omega)$ (Figs.~\ref{fig.decomposition}(b) and~\ref{fig.decompositiona}(b)), which leads to good agreement for $M_{N}^{\omega < \omega^\ast}$.
The discrepancy in $M_{N}^{\omega < \omega^\ast}$ of the stressed system can be adjusted by tuning the exponents of $a$ and $b$ to take into account the smooth crossovers.
In Fig.~\ref{fig.nonaffine}(a), we also plot Eq.~(\ref{nonaffine2}) with $a=1.5$ and $b=1.3$ (dashed lines);
\begin{equation} \label{nonaffine2x}
M_{N}^{\omega < \omega^\ast} = \left(2.5 \right) \hat{\rho}_c g^\ast M_N^{\ast} {\omega^\ast} + \mathcal{O}(\Delta \varphi) \quad \text{(stressed)},
\end{equation}
which works better to capture the simulation values.

The total modulus, $M_N=M_{N}^{\omega < \omega^\ast}+M_{N}^{\omega > \omega^\ast}$, is then acquired by Eq~(\ref{nonaffine9}), as demonstrated in Fig.~\ref{fig.nonaffine}(e),(f).
Again, for the stressed system in (e), the dashed line plots Eq~(\ref{nonaffine9}) with $a=1.5$ and $b=1.3$;
\begin{equation} \label{nonaffine9x}
M_{N} = M_{Nc} - \left(0.5 \right) \hat{\rho}_c g^\ast M_{N}^{\ast} \omega^\ast + \mathcal{O}(\Delta \varphi) \quad \text{(stressed)}.
\end{equation}
On approach to the transition point $\varphi_c$, the frequency $\omega^\ast$ goes to zero, hence the non-affine modulus $M_N$ tends towards the critical value $M_{Nc}$, as $M_N-M_{Nc} \sim \omega^\ast \sim \Delta \varphi^{1/2} \rightarrow 0$.
We note that the critical value of $M_{Nc}$ is a finite positive value (see Eq.~(\ref{nonaffine7})), like the affine modulus $M_{Ac}$ in Eq.~(\ref{affine5a}), thus $M_N$ also discontinuously goes to zero, through the transition to the fluid phase, $\varphi < \varphi_c$, where $M_N \equiv 0$.

\begin{figure}[t]
\centering
\includegraphics[width=0.48\textwidth]{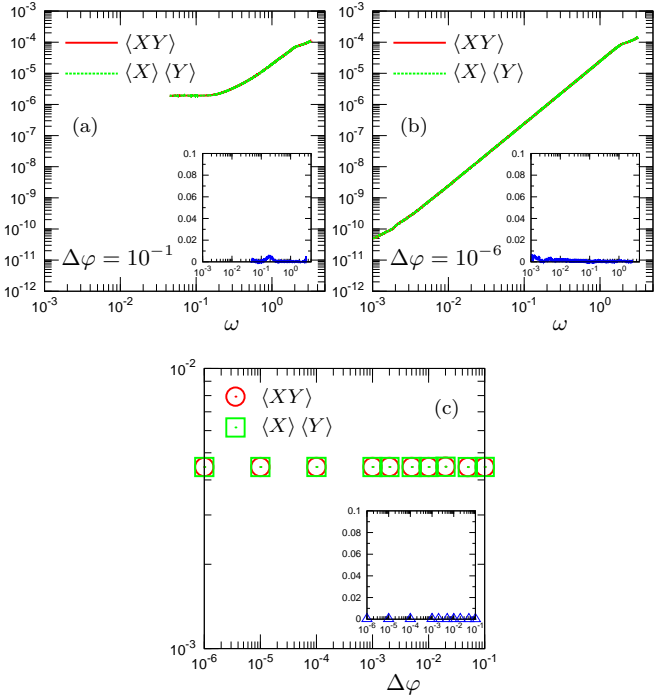}
\vspace*{0mm}
\caption{\label{fig.correlation}
(Color online) Correlation between two quantities $X$ and $Y$; $X=( n^x_{ij} n^y_{ij} )^2$, $Y=(\vec{e}^k_{ij}\cdot \vec{n}_{ij} )^2$ in (a),(b) and $X=( n^x_{ij} n^y_{ij} )^2$, $Y=(n^x_{i'j'} n^y_{i'j'} )^2$ in (c).
In the main panel, we plot $\left< XY \right>$ and $\left< X \right> \left< Y \right>$, as a function of the eigenfrequency $\omega$ in (a),(b) and the packing fraction $\Delta \varphi$ in (c).
In (a),(b), the values are averaged over frequency bins of $\log_{10} \omega^k \in [\log_{10} \omega-\Delta \omega/2,\log_{10} \omega + \Delta \omega/2]$ with $\Delta \omega = 0.07$, and the packing fraction is (a) $\Delta \varphi = 10^{-1}$ and (b) $\Delta \varphi = 10^{-6}$.
If $X$ and $Y$ are uncorrelated, $\left< XY \right> = \left< X \right> \left< Y \right>$ holds.
To see this quantitatively, we plot the relative error, $\left| \left< XY \right> - \left< X \right> \left< Y \right> \right|/\left| \left< X \right> \left< Y \right> \right|$, in the insets.
We observe non-correlations (zero correlations) between $X$ and $Y$, in all the cases of (a), (b), and (c).}
\end{figure}

\subsection{Critical values of elastic moduli at the transition} \label{sec.critical}
Until now, we have shown that the affine modulus $M_A$ approaches the critical value $M_{Ac}$ as the excess contact number $\Delta z \sim \Delta \varphi^{1/2}$ vanishes, while the non-affine modulus $M_N$ likewise goes to $M_{Nc}$ as the crossover frequency $\omega^\ast \sim \Delta \varphi^{1/2}$ goes to zero.
It is worth noting that $\Delta z$ and $\omega^\ast$ have the same power-law exponent $1/2$ with respect to $\Delta \varphi$; $\Delta z \sim \omega^\ast \sim \Delta \varphi^{1/2}$~\cite{Silbert_2005,Wyart_2005,Wyart_2006}.
The behaviors of the affine, $M_A$, and non-affine, $M_N$, moduli are similar between the bulk $M_{A,N}=K_{A,N}$ and the shear $M_{A,N}=G_{A,N}$ moduli.
However, the total moduli, $K=K_A-K_N$ and $G=G_A-G_N$, show distinct critical behaviors through the transition $\varphi_c$ to the fluid phase~\cite{OHern_2002,OHern_2003,Ellenbroek_2006,Ellenbroek_2009,Ellenbroek2_2009,Wyart}: The total bulk modulus $K$ discontinuously drops to zero, while the total shear modulus $G$ continuously goes to zero, which are described by the power-law scalings, $K \sim \Delta \varphi^0$ and $G \sim \Delta \varphi^{1/2}$ in Eq.~(\ref{plmacro}) and Fig.~\ref{fig.phidependence}(a),(b).
This difference is due to the distinct critical values of $K_c=K_{Ac}-K_{Nc}$ and $G_c=G_{Ac}-G_{Nc}$ at the transition $\varphi_c$.
$K_{Ac}$ is larger than $K_{Nc}$, $K_{Ac}\simeq 0.40 >K_{Nc} \simeq 0.15$, leading to a finite value of $K_c=0.25$.
On the other hand, $G_{Ac}$ and $G_{Nc}$ coincide, $G_{Ac}=G_{Nc} \simeq 0.24$, resulting in zero total shear modulus $G_c=0$.
Our final goal in this section is to derive these critical values, using Eq.~(\ref{modulus2}) for $K_{Ac}, G_{Ac}$, and Eq.~(\ref{modulus11}) for $K_{Nc}, G_{Nc}$.

{\bf Critical values of affine moduli $\boldsymbol{M_{Ac}}$.}
At the transition point $\varphi_c$, the system is in the isostatic state~\cite{Wyart,Wyart_2005,Wyart_2006,Maxwell_1864}, where the number of contacts precisely equals the degrees of freedom $3N-3$;
\begin{equation}
N^\text{ct}_c=3N-3 \left( =\frac{N z_c}{2} \right).
\end{equation}
In addition, since the pressure is zero, $p=0$, there should be no overlaps at all the particle contacts $(i,j)$, i.e.,
\begin{equation}
\vec{r}_{ij} \equiv \vec{n}_{ij}, \qquad r_{ij} \equiv 1, \qquad \phi'(r_{ij}) \equiv 0,
\end{equation}
hold for all $N^\text{ct}_c$ contacts $(i,j)$.
Note that at $\varphi_c$, the stressed and unstressed systems are exactly same.
We therefore use Eq.~(\ref{modulus2}) to evaluate the critical values $K_{Ac}, G_{Ac}$ as
\begin{equation}
\begin{aligned} \label{loss1}
K_{Ac} &= \frac{1}{V} \sum_{ (i,j) \in N^\text{ct}_c } \frac{1}{9} = \frac{ N^\text{ct}_c }{9V},\\
G_{Ac} &= \frac{1}{V} \sum_{ (i,j) \in N^\text{ct}_c } \left( {n_{ij}^{x}} {n_{ij}^{y}} \right)^2 = \frac{N^\text{ct}_c}{V} \left< \left( {n_{ij}^{x}} {n_{ij}^{y}} \right)^2 \right>,
\end{aligned}
\end{equation}
where $\left< \right>$ denotes the average value over all of $N^\text{ct}_c$ contacts.
$K_{Ac}$ is exactly the same as that in Eq.~(\ref{affine5a}).
Also, the isotropic distribution of the bond vector $\vec{n}_{ij}$, Eq.~(\ref{affine2}), recovers $G_{Ac}$ in Eq.~(\ref{affine5a}), as done in Sec.~\ref{sec.affine}.

{\bf Critical values of non-affine moduli $\boldsymbol{M_{Nc}}$.}
We next formulate $K_{Nc}, G_{Nc}$ from Eq.~(\ref{modulus11}).
The bulk modulus $K_{Nc}$ is formulated as
\begin{equation}
\begin{aligned} \label{loss3a}
& K_{Nc} = \frac{1}{V} \sum_{k=1}^{3N-3} \frac{1}{{\omega^k}^2} \left[ V \sum_{(i,j)} \frac{\partial p}{\partial \vec{r}_{ij}} \cdot \vec{e}^k_{ij} \right]^2, \\
& = \frac{1}{V} \sum_{k=1}^{3N-3} \frac{1}{{\omega^k}^2} \left[ \sum_{(i,j)} \frac{1}{3} \left( \vec{e}^k_{ij}\cdot \vec{n}_{ij} \right) \right]^2, \\
& = \frac{1}{9V} \sum_{k=1}^{3N-3} \frac{1}{{\omega^k}^2} \sum_{(i,j)} \left( \vec{e}^k_{ij}\cdot \vec{n}_{ij} \right)^2 \\
& + \frac{1}{9V} \sum_{k=1}^{3N-3} \frac{1}{{\omega^k}^2} \sum_{(i,j)} \sum_{(i',j') \neq (i,j)} \left( \vec{e}^k_{ij}\cdot \vec{n}_{ij} \right) \left( \vec{e}^k_{i'j'}\cdot \vec{n}_{i'j'} \right),\\
& = K_{Ac} \\
& + \frac{N^\text{ct}_c(N^\text{ct}_c-1)}{9V} \left[ \sum_{k=1}^{3N-3} \frac{\left< \left( \vec{e}^k_{ij}\cdot \vec{n}_{ij} \right) \left( \vec{e}^k_{i'j'}\cdot \vec{n}_{i'j'} \right) \right>}{{\omega^k}^2} \right].
\end{aligned}
\end{equation}
In the derivation of Eq.~(\ref{loss3a}), we use Eq.~(\ref{vs4}) at the transition point $\varphi_c$, i.e.,
\begin{equation} \label{loss2a}
\sum_{(i,j) \in N^\text{ct}_c} \left( \vec{e}^k_{ij}\cdot \vec{n}_{ij} \right)^2 = {\omega^k}^2.
\end{equation}

To formulate the shear modulus $G_{Nc}$, we assume that (i) ${n_{ij}^{x}} {n_{ij}^{y}}$ and $\left( \vec{e}^k_{ij} \cdot \vec{n}_{ij} \right)$ are uncorrelated in each mode $k$;
\begin{equation} \label{lossassume1}
\left< \left( {n_{ij}^{x}} {n_{ij}^{y}} \right) \left( \vec{e}^k_{ij}\cdot \vec{n}_{ij} \right) \right> = \left< {n_{ij}^{x}} {n_{ij}^{y}} \right> \left< \vec{e}^k_{ij}\cdot \vec{n}_{ij} \right>,
\end{equation}
and (ii) ${n_{ij}^{x}} {n_{ij}^{y}}$ and ${n_{i'j'}^{x}} {n_{i'j'}^{y}}$ at different contacts, $(i,j) \neq (i',j')$, are also uncorrelated;
\begin{equation} \label{lossassume2}
\left< \left({n_{ij}^{x}} {n_{ij}^{y}} \right) \left({n_{i'j'}^{x}} {n_{i'j'}^{y}} \right) \right> = \left< {n_{ij}^{x}} {n_{ij}^{y}} \right>^2.
\end{equation}
%
Those two assumptions are numerically verified by Fig.~\ref{fig.correlation}, for (i) in (a),(b) and (ii) in (c), where for convenience, we study correlations of the quantities $\left( n_{ij}^{x} n_{ij}^{y} \right)^2$ and $\left( \vec{e}^k_{ij} \cdot \vec{n}_{ij} \right)^2$, instead of $n_{ij}^{x} n_{ij}^{y}$ and $\vec{e}^k_{ij} \cdot \vec{n}_{ij}$.
We have also confirmed that the assumptions (i) and (ii) hold for the range of packing fraction, $\Delta \varphi = 10^{-1}$ to $10^{-6}$.
Using Eqs.~(\ref{lossassume1}) and~(\ref{lossassume2}), we can formulate the shear modulus $G_{Nc}$ as
\begin{equation}
\begin{aligned} \label{loss3b}
& G_{Nc} = \frac{1}{V} \sum_{k=1}^{3N-3} \frac{1}{{\omega^k}^2} \left[ V \sum_{(i,j)} \frac{\partial \sigma_s}{\partial \vec{r}_{ij}} \cdot \vec{e}^k_{ij} \right]^2, \\
& = \frac{1}{V} \sum_{k=1}^{3N-3} \frac{1}{{\omega^k}^2} \left[ \sum_{(i,j)} {n_{ij}^{x}} {n_{ij}^{y}} \left( \vec{e}^k_{ij} \cdot  \vec{n}_{ij} \right) \right]^2, \\
& = \frac{1}{V} \sum_{k=1}^{3N-3} \frac{1}{{\omega^k}^2} \sum_{(i,j)} \left( {n_{ij}^{x}} {n_{ij}^{y}} \right)^2 \left( \vec{e}^k_{ij}\cdot \vec{n}_{ij} \right)^2 \\
& + \frac{1}{V} \sum_{k=1}^{3N-3} \frac{1}{{\omega^k}^2} \sum_{(i,j)} \sum_{(i',j') \neq (i,j)} {n_{ij}^{x}} {n_{ij}^{y}} {n_{i'j'}^{x}} {n_{i'j'}^{y}} \\
& \times \left( \vec{e}^k_{ij}\cdot \vec{n}_{ij} \right) \left( \vec{e}^k_{i'j'}\cdot \vec{n}_{i'j'} \right), \\
& = \frac{N^\text{ct}_c}{V} \sum_{k=1}^{3N-3} \frac{1}{{\omega^k}^2} \left< \left( {n_{ij}^{x}} {n_{ij}^{y}} \right)^2 \right> \left<\left( \vec{e}^k_{ij}\cdot \vec{n}_{ij} \right)^2 \right> \\
& + \frac{N^\text{ct}_c(N^\text{ct}_c-1)}{V} \sum_{k=1}^{3N-3} \frac{1}{{\omega^k}^2} \left< {n_{ij}^{x}} {n_{ij}^{y}} {n_{i'j'}^{x}} {n_{i'j'}^{y}} \right> \\
& \times \left< \left( \vec{e}^k_{ij}\cdot \vec{n}_{ij} \right) \left( \vec{e}^k_{i'j'}\cdot \vec{n}_{i'j'} \right) \right>, \\
& = \frac{N^\text{ct}_c}{V} \left< \left( {n_{ij}^{x}} {n_{ij}^{y}} \right)^2 \right> + \left< {n_{ij}^{x}} {n_{ij}^{y}} \right>^2 \\
& \times \left\{ \frac{N^\text{ct}_c (N^\text{ct}_c-1)}{V} \left[ \sum_{k=1}^{3N-3} \frac{\left< \left( \vec{e}^k_{ij}\cdot \vec{n}_{ij} \right) \left( \vec{e}^k_{i'j'}\cdot \vec{n}_{i'j'} \right) \right>}{{\omega^k}^2} \right] \right\},\\
& = G_{Ac}.
\end{aligned}
\end{equation}
In the final equality of Eq.~(\ref{loss3b}), we use $\left< {n_{ij}^{x}} {n_{ij}^{y}} \right> = 0$, which is obtained by the isotropic distribution of $\vec{n}_{ij}$, Eq.~(\ref{affine2}).
Therefore, the non-affine value $G_{Nc}$ \textit{exactly} coincides with the affine value $G_{Ac}$.

{\bf Critical values of total moduli $\boldsymbol{M_{c}}$.}
From Eqs.~(\ref{loss3a}) and~(\ref{loss3b}), we obtain
\begin{equation}
\begin{aligned} \label{loss4a}
& K_c = K_{Ac} - K_{Nc}, \\
&= - \frac{N^\text{ct}_c(N^\text{ct}_c-1)}{9V} \left[ \sum_{k=1}^{3N-3} \frac{\left< \left( \vec{e}^k_{ij}\cdot \vec{n}_{ij} \right) \left( \vec{e}^k_{i'j'}\cdot \vec{n}_{i'j'} \right) \right>}{{\omega^k}^2} \right], \\
& G_c = G_{Ac} - G_{Nc}, \\
&= \left< {n_{ij}^{x}} {n_{ij}^{y}} \right>^2 \times \left(9 K_c \right) = 0.
\end{aligned}
\end{equation}
The finite value of the bulk modulus $K_c$ is given by the correlations of the angle of vibrational motion relative to bond vector, between different contacts $(i,j) \neq (i',j')$, $\left< \left( \vec{e}^k_{ij} \cdot \vec{n}_{ij} \right) \left( \vec{e}^k_{i'j'} \cdot \vec{n}_{i'j'} \right) \right>$.
We numerically get
\begin{equation} \label{loss5}
\left[ \sum_{k=1}^{3N-3} \frac{\left< \left( \vec{e}^k_{ij}\cdot \vec{n}_{ij} \right) \left( \vec{e}^k_{i'j'}\cdot \vec{n}_{i'j'} \right) \right>}{{\omega^k}^2} \right] = -2.1\times 10^{-4},
\end{equation}
which confirms the value of $K_c=\left[ {N^\text{ct}_c(N^\text{ct}_c-1)}/{9V} \right] \times \left( 2.1\times 10^{-4} \right) \simeq 0.25$.
For the shear modulus $G_c$, the correlation term disappears due to the term, $\left< {n_{ij}^{x}} {n_{ij}^{y}} \right> = 0$, giving the zero value of $G_c=0$.
The zero shear modulus $G_c$ is based on two features of jammed solids: (i) The bond vector $\vec{n}_{ij}$ and the contact vibration $\vec{e}_{ij} \cdot \vec{n}_{ij}$ are uncorrelated (see Eq.~(\ref{lossassume1})), and (ii) the bond vector $\vec{n}_{ij}$ is randomly and isotropically distributed (see Eqs.~(\ref{affine2}) and~(\ref{lossassume2})).
Thus, it is those two features, (i) and (ii), that cause the distinction between the critical values and behaviors of the bulk $K$ and the shear $G$ moduli, in marginally jammed
solids.
Interestingly, Zaccone and Terentjev~\cite{Zaccone_2014} have theoretically explained the finite value of bulk modulus $K_c$ by taking into account the excluded-volume correlations between different contacts, $(i,j) \neq (i',j')$.
They also demonstrated that the excluded-volume correlations are weaker under shear, leading to a smaller value of shear modulus $G_c$.
The correlation term, $\left< \left( \vec{e}^k_{ij} \cdot \vec{n}_{ij} \right) \left( \vec{e}^k_{i'j'} \cdot \vec{n}_{i'j'} \right) \right>$, in Eq.~(\ref{loss4a}) may be related to such excluded-volume correlations.

\section{Summary} \label{summary}
{\bf Scaling behaviors with $\boldsymbol{\Delta z}$, $\boldsymbol{\omega^\ast}$, and $\boldsymbol{\Delta \varphi}$.}
In the present paper, using the harmonic formulation~\cite{Lutsko_1989,Maloney_2004,Maloney2_2006,Maloney_2006,Karmakar_2010,Lemaitre_2006,Hentschel_2011,Zaccone_2011,Zaccone2_2011,Zaccone_2014}, we have studied the elastic moduli $M=K,G$ in a model jammed solid for a linear interaction force law, close to the (un)jamming transition point $\varphi_c$.
As we approach the transition point $\varphi_c$, $\Delta \varphi \rightarrow 0$, the excess contact number goes to zero, $\Delta z \rightarrow 0$, and at the same time, vibrational eigenmodes in the plateau regime of $g(\omega)$ extend towards zero frequency, $\omega^\ast \rightarrow 0$.
Accordingly, the affine modulus, $M_A=K_A,G_A$, tends towards the critical value, $M_{Ac}=K_{Ac},G_{Ac}$, as $M_A - M_{Ac} \sim \Delta z \rightarrow 0$ (Eqs.~(\ref{affine5x}), (\ref{affine5}), Fig.~\ref{fig.affine}), whereas the non-affine modulus, $M_N=K_N,G_N$, converges to $M_{Nc}=K_{Nc},G_{Nc}$, following $M_N - M_{Nc} \sim \omega^\ast \rightarrow 0$ (Eqs.~(\ref{nonaffine2}), (\ref{nonaffine6}), (\ref{nonaffine9}), Fig.~\ref{fig.nonaffine}).
Thus, the total modulus, $M = M_A-M_N$, is
\begin{equation} \label{rconclusion1}
M = M_{c} + \alpha_M \Delta z - \beta_M \omega^\ast = M_{c} + \gamma_M \Delta \varphi^{1/2},
\end{equation}
where $M_c=M_{Ac}-M_{Nc}$ is the critical value of $M$, and $\alpha_M, \beta_M, \gamma_M$ are coefficients.
As numerically~\cite{Silbert_2005} and theoretically~\cite{Wyart_2005,Wyart_2006} demonstrated, $\Delta z$ and $\omega^\ast$ have the same power-law scalings with $\Delta \varphi$, i.e., $\Delta z \sim \omega^\ast \sim \Delta \varphi^{1/2}$, which gives the second equality in Eq.~(\ref{rconclusion1}), and $M-M_c \sim \Delta z \sim \Delta \varphi^{1/2}$.

{\bf Origin of distinct critical values between bulk and shear moduli.}
Both the bulk, $M=K$, and shear, $M=G$, moduli share the same behavior of Eq.~(\ref{rconclusion1}), but, crucially, a difference between the two elastic moduli appears in their critical values, $K_c, G_c$.
For the bulk modulus, $K_{Ac}$ is larger than $K_{Nc}$, and the total value $K_c$ is a finite, positive constant.
In contrast, $G_{Ac}$ and $G_{Nc}$, exactly match, and the total shear modulus $G_c$ is zero.
This difference causes distinct critical behaviors: $K = K_c + \gamma_K \Delta \varphi^{1/2} \sim \Delta \varphi^{0}$ and $G = \gamma_G \Delta \varphi^{1/2} \sim \Delta \varphi^{1/2}$ (Eq.~(\ref{plmacro}), Fig.~\ref{fig.phidependence}).
Thus, through the unjamming transition into the fluid phase ($\varphi < \varphi_c $), $K$ discontinuously drops to zero, whereas $G$ continuously vanishes \cite{OHern_2002,OHern_2003,Ellenbroek_2006,Ellenbroek_2009,Ellenbroek2_2009,Wyart}.
In the present work, we showed that the finite bulk modulus $K_c$ is controlled by correlations between contact vibrational motions, $\vec{e}^k_{ij} \cdot \vec{n}_{ij}$ and $\vec{e}^k_{i'j'} \cdot \vec{n}_{i'j'}$, at different contacts $(i,j) \neq (i',j')$ (Eq.~(\ref{loss4a})), which might be related to excluded-volume correlations as suggested by Zaccone and Terentjev~\cite{Zaccone_2014}.
In the case of the shear modulus $G_c$, such correlations are washed out by two key features of jammed, disordered solids: (i) The contact bond $\vec{n}_{ij}$ and the contact vibrational motions $\vec{e}^k_{ij} \cdot \vec{n}_{ij}$ are uncorrelated, and (ii) the contact bond $\vec{n}_{ij}$ is randomly and isotropically distributed (Eqs.~(\ref{affine2}), (\ref{lossassume1}), (\ref{lossassume2}), Figs.~\ref{fig.angle}, \ref{fig.correlation}).
In the end, the critical value $G_c$ becomes \textit{exactly} zero (Eq.~(\ref{loss4a})).

{\bf Eigenmode decomposition of non-affine moduli $\boldsymbol{M_N}$.}
A main result of the present work is the eigenmode decomposition of the non-affine elastic moduli $M_N$, as presented in Figs.~\ref{fig.decomposition} and~\ref{fig.decompositiona} for $M_N=K_N$ and $G_N$, respectively.
The modal contribution to the non-affine modulus, $M^k_N$, shows three distinct regimes in frequency $\omega$ space, with two crossovers at $\omega =\omega^\ast$ and $\omega = \omega^h$, which match precisely the regimes already apparent in the vDOS $g(\omega)$ (Figs.~\ref{fig.vibration} and~\ref{fig.frequency}).
We showed that the crossover point $\omega^\ast$ is controlled by the competition between two vibrational energies, the compressing/stretching energy, $\delta E^{k \parallel}$, and the sliding energy, $\delta E^{k \perp}$, whereas the crossover at $\omega^h$ is determined by the balance between two vibrational motions along the bond vector $\vec{n}_{ij}$, compressing motion, ${e}^{k \parallel}_\text{com}$, and stretching motion, ${e}^{k \parallel}_\text{str}$.

The behavior of $M^k_N = \left| \Sigma_M^k \right| \times \left| \delta {R}_{\text{na} M}^k \right|$ (dependence of $M^k_N$ on $\omega$) is understood in terms of the energy relaxation during non-affine deformation process.
During the non-affine deformation, high-frequency modes with $\omega > \omega^h$ are only weakly activated, leading to a relatively small contribution to the non-affine modulus.
At intermediate frequencies, $\omega^\ast < \omega < \omega^h$, modes of lower energy are more readily activated, which increases $\left| \delta {R}_{\text{na} M}^k \right| \sim \omega^{-1}$ and thereby enhances $M_N^k$.
However the lower $\omega$ modes also generate smaller forcings, $\left| \Sigma_M^k \right| \sim \omega$, reducing $M_N^k$ with decreasing frequency.
These two opposite $\omega$-dependences of $\left| \Sigma_M^k \right|$ and $\left| \delta {R}_{\text{na} M}^k \right|$ lead to the frequency-independent behavior of $M^k_N$, as $M^k_N = \left| \Sigma_M^k \right| \times \left| \delta {R}_{\text{na} M}^k \right| \sim \omega^{0}$.
Finally at the lower end of the spectrum, $\omega < \omega^\ast$, for the stressed system, the stress, $\sim \phi'(r_{ij}) \sim \Delta \varphi$, enhances the force $\left| \Sigma_M^k \right| \sim \omega^0$ and drives the non-affine displacement $\left| \delta {R}_{\text{na} M}^k \right| \sim \omega^{-2}$.
As a result, the energy relaxation $M_N^k$ grows with decreasing $\omega$ as $M^k_N \sim \omega^{-2}$.
Such effects are not observed for the unstressed system, with zero stress, $\phi'(r_{ij}) \equiv 0$, i.e., the unstressed system retains the frequency-independent behavior, $M^k_N \sim \omega^{0}$.
In all the cases, the above behaviors of $M_N^k$ (and $\left| \Sigma_M^k \right|,\left| \delta {R}_{\text{na} M}^k \right|$) are controlled by the net compressional/stretching motions $e^{k \parallel}_\text{net}$ (Eqs.~(\ref{resultvs2}),(\ref{resultenedis}), Figs.~\ref{fig.vibration}, \ref{fig.decomposition}, \ref{fig.decompositiona}).

{\bf Non-affine motions and low frequency mode excitations.}
Large-scale, non-affine motions of particles have been reported for athermal jammed solids~\cite{Maloney_2004,Maloney2_2006,Maloney_2006,Lemaitre_2006}, and also for thermal glasses~\cite{Wittmer_2002,Tanguy_2002,leonforte_2005,DiDonna_2005}.
Our results indicate that such large-scale non-affine displacement fields are induced through the low-frequency eigenmodes excitations (Eq.~(\ref{resultenedis}), Figs.~\ref{fig.decomposition}, \ref{fig.decompositiona}, and~\ref{fig.picture}): On approach to the transition point $\varphi_c$, lower frequency modes $k$ are more readily activated, resulting in larger non-affine displacements $\left| \delta {R}_{\text{na} M}^k \right|$.
Since the lower frequency modes exhibit more floppy-like vibrational motions (Fig.~\ref{fig.alpha}), the non-affine motions correspondingly exhibit floppy-like character closer to $\varphi_c$, which is consistent with previous works~\cite{Ellenbroek_2006,Ellenbroek_2009,Ellenbroek2_2009}.

As reported in Refs.~\cite{Ellenbroek_2006,Ellenbroek_2009,Ellenbroek2_2009}, the floppy-like, non-affine motions are more prominent under shear deformation than under compression, which thereby makes a distinction between these two elastic responses.
Thus, at first sight, it seems natural to associate the low-frequency, floppy-like modes as being wholly responsible for such the distinction between compression and shear.
However, we have shown that the difference in the nonaffine responses between compression and shear is largely controlled by relatively high-frequency eigenmodes with $\omega \gtrsim 10^{-1}$, not solely by low frequency modes (Fig.~\ref{fig.decomposition2}).
Low frequency mode excitations for $\omega \lesssim 10^{-1}$ are very similar between bulk and shear deformations, while it is those modes with $\omega \gtrsim 10^{-1}$ that are more readily activated under shear than under compression.
Thus, the mode excitations at $\omega \gtrsim 10^{-1}$ contribute significantly to the non-affine shear modulus $G_N$, causing it to become enhanced over the bulk modulus $K_N$.
Ultimately, the critical value $G_{Nc} \simeq 0.24$ is larger than $K_{Nc} \simeq 0.15$.

\section{Conclusions and perspectives} \label{conclusions}
{\bf Characterization of the frequency $\boldsymbol{\omega^\ast}$.}
The vibrational modes are directly related to the elastic properties (non-affine elastic moduli) of the system.
In the case of the marginally jammed packings studied here, the modal contribution $M^k_N$ to the non-affine elastic modulus shows a frequency-independent plateau, $M^k_N \sim \omega^0$, above the frequency $\omega^\ast$.
This characteristic feature is attributed to the fact that only compressing/stretching vibrational motions contribute to the mode energy, whereas the sliding vibrations feel few constraints, making a negligible contribution to the vibrational energy.
However, below $\omega^\ast$, sliding motions play a role in the total mode energy, causing the crossover behavior of $M^k_N$ at $\omega^\ast$, from $\sim \omega^0$ to $\sim \omega^{-2}$.
Wyart~\textit{et. al.}~\cite{Wyart_2005,Wyart_2006,Xu_2007,Goodrich_2013} have characterized the frequency $\omega^\ast$ in terms of a purely geometric property (variational arguments), where the excess contact number $\Delta z$ controls $\omega^\ast$.
In addition, the energy diffusivity in heat transport as well as the dynamical structure factor show crossover behaviors at $\omega^\ast$ in the \textit{unstressed} system~\cite{Xu_2009,Vitelli_2010}, which have been then theoretically described using the effective medium approach~\cite{Wyart_2010,DeGiuli_2014}.
In the present work, we have marked $\omega^\ast$ as the characteristic frequency where the two vibrational energies, the compressing/stretching and the sliding vibrational energies, become comparable to each other in the \textit{stressed} system, which induces a crossover in the energy-related quantities including the elastic modulus $M^k_N$.

{\bf Debye regime in vDOS and continuum limit.}
As shown in Fig.~\ref{fig.vibration}(a),(b) of vDOS $g(\omega)$, we do not observe the expected Debye scaling regime, $g(\omega) \sim \omega^{2}$, in the low frequency limit.
We have also confirmed that even the lowest eigenmodes in our frequency window are far from plane-wave modes which are also expected to appear at low frequencies.
The Debye scaling and the plane-wave modes are likely to be observed by employing larger system sizes to access lower frequencies.
Yet, this aspect of the vDOS remains an open issue for jammed particulate systems whereby the so-called Boson peak appears to extend down to zero frequency $\omega = 0$ as $\Delta\varphi \to 0$.
We might expect that the deviations from traditional Debye scaling in the low-$\omega$ tail of the vDOS are generic to amorphous materials and also tunable through packing structure~\cite{Schreck2_2011,Mizuno3_2016}.

In a different, yet related, context, recent numerical works~\cite{Ellenbroek_2006,Ellenbroek_2009,Lerner_2014,Karimi_2015} have discussed the continuum limit by studying the mechanical response to \textit{local} forcing.
This continuum limit corresponds to a scale above which the elastic properties match those of the entire, bulk system.
Whereas below this length scale the elastic response differs from the bulk average, and local elasticity becomes apparent.
At the low frequencies corresponding to wavelengths comparable to this continuum limit, we might expect the vibrational modes to be compatible with the plane-wave modes described by continuum mechanics.
Although here we caution that the length scale at which a local elastic description coincides with bulk behavior diverges as $\Delta \varphi \to 0$~\cite{Ellenbroek_2006,Ellenbroek_2009,Lerner_2014,Karimi_2015}.

{\bf System size effects on elastic moduli values.}
In the present work, we have employed relatively small systems with $N \approx 1000$ ($L \approx 9$).
As mentioned above, we do not access very low frequency modes where one might expect to observe the Debye scaling regime in the vDOS.
We therefore consider that the lack of lower frequency modes may cause some finite system size effects in the non-affine elastic moduli values, $M_N$.
Indeed, Ref.~\cite{Tanguy_2002} has reported system size effects appearing in two-dimensional Lennard-Jones glasses with small system sizes.
For the present jammed systems, recent numerical work~\cite{Goodrich_2012} calculated the elastic moduli, changing the system size from $N=64$ to $4096$.
In the results of Ref.~\cite{Goodrich_2012}, for our studied pressure regime, we do not find any noticeable differences in the elastic moduli values between different system sizes of $N \gtrsim 1000$.
Particularly, the scaling laws with packing fraction $\Delta \varphi$ are consistent for all the system sizes of $N \gtrsim 1000$.
Also, we have confirmed that our values of the elastic moduli and scaling laws with $\Delta \varphi$ are consistent with the values of $N \gtrsim 1000$ in Ref.~\cite{Goodrich_2012}.
This observation indicates that our moduli values are not influenced by system size effects.
Thus, we conclude that for system sizes $N \gtrsim 1000$, the lack of accessing lower frequency modes, including those in the Debye regime, does not qualitatively impact our results for the elastic moduli, and therefore, does not change the scaling laws with $\Delta \varphi$.
In order to demonstrate this conclusion more explicitly, it could an interesting future work to measure the modal contribution of $M^k_N$ in the Debye regime, using large systems.

{\bf Effects of friction, particle-size ratio, particle shape, and deeply jammed state.}
It has been reported that jammed packings, composed of frictional particles~\cite{Somfai_2007,Silbert_2010,Still_2014}, mixtures with large particle-size ratio~\cite{Xu2_2010}, and non-spherical particles (e.g., ellipse-shaped particles)~\cite{Zeravcic_2009,Mailman_2009}, show some distinct features in the vibrational and mechanical properties, from those of the frictionless sphere packings studied in the present work.
Effects of friction, particle-size ratio, and particle shape on the mechanical properties are a timely subject.
The modal decomposition of the non-affine moduli allows us to connect unusual features apparent in the vibrational spectrum to the elastic moduli properties, as we have performed here on the sphere packings.
Another interesting study could be on deeply jammed systems at very high packing fractions~\cite{Zhao_2011,Ciamarra_2013}.
Deeply jammed systems show anomalous vibrational and mechanical properties, particularly different power-law scalings from those of the marginally jammed solids~\cite{Zhao_2011}.
In addition, high-order jamming transitions accompanying the mechanical and density anomalies have been reported~\cite{Ciamarra_2013}.
It would be an interesting subject to explore the role of vibrational anomalies on the mechanical properties of such systems.

{\bf Local elastic moduli distribution, soft spot, and low frequency modes.}
Amorphous materials exhibit spatially heterogeneous distributions of local elastic moduli, as has been demonstrated by simulations~\cite{Mayr_2009,tsamados_2009,Mizuno_2013,yoshimoto_2004,makke_2011,Mizuno2_2013,Mizuno_2014} and experiments~\cite{Wagner_2011}.
Recent numerical works~\cite{Mizuno_2016,Cakir_2016} have studied the local elastic moduli distributions in jammed packings.
Manning and co-workers~\cite{Manning_2011,chen_2011} proposed that ``soft spots'' can be associated with regions of atypically large displacements of particles in the quasi-localized, low-frequency vibrational modes.
It has been reported that particle rearrangements, which are activated by mechanical load~\cite{Manning_2011,Tanguy_2010} and by thermal energy~\cite{chen_2011,widmer_2008}, tend to occur in those so-called soft spots.
Thus, we could assume that the soft spots, which are detected by the low frequency (localized) modes, are linked to the low elastic moduli regions.
In the present work, we demonstrated that the non-affine elastic modulus is determined mainly by the vibrational modes excitations at $\omega >\omega^\ast$, whereas the low frequency modes with $\omega < \omega^\ast$ make only small contributions to elastic moduli.
Our result therefore indicates that the low frequency modes themselves do not influence the elastic properties, but rather they are just driven by the elastic moduli distributions constructed by the modes with $\omega > \omega^\ast$.

In the case of marginally jammed solids, the shear modulus becomes orders of magnitude smaller than the bulk modulus (see Fig.~\ref{fig.phidependence}), thus the low frequency modes are most likely related to the shear modulus.
In addition, in our study~\cite{Mizuno_2016}, we have demonstrated that spatial fluctuations of the local shear modulus grow on approach to the jamming transition $\varphi_c$.
Therefore, those observations could assume that the growing \textit{shear} modulus heterogeneities drive the low frequency modes excitations, particularly the localizations of low frequency modes.
Schirmacher \textit{et. al.}~\cite{Marruzzo_2013,Schirmacher_2015,Schirmacher2_2015} have constructed such a heterogeneous-elasticity theory based on this picture, where the shear modulus heterogeneities determine the behavior of low frequency modes, e.g., the Boson peak.
Those topics, focusing on the local elastic moduli, soft spots, and low frequency modes, could be an important future work.

{\bf Generalization to other contacts, and non-linear effects.}
We have studied the linear elastic properties throughout the present paper.
As long as we stay in the linear elastic regime, our results, which have been obtained from the harmonic potential, can be extended to other potentials;
\begin{equation}
\phi(r_{ij}) =
\left\{ \begin{aligned}
& \frac{\varepsilon}{a} \left(1-\frac{r_{ij}}{\sigma} \right)^a & (r_{ij} < \sigma), \\
& 0 & (r_{ij} \ge \sigma),
\end{aligned} \right. \\
\end{equation}
where $a>0$ characterizes the potential, e.g., $a=2$ is the present harmonic potential case, while $a=2.5$ provides Hertzian contacts, by considering ``normalized variables", e.g., normalized elastic modulus and frequency;
\begin{equation}
\begin{aligned}
\hat{M}= \frac{M}{k_\text{eff}}, \qquad \hat{\omega} = \frac{\omega}{k_\text{eff}^{1/2}},
\end{aligned}
\end{equation}
where the values are normalized by the effective spring constant, $k_\text{eff} \sim \phi'' \sim \Delta \varphi^{a-2}$~\cite{Vitelli_2010}.
However, marginally jammed solids are highly sensitive to non-linear effects caused by thermal agitation or finite large strain, as actively discussed in recent works~\cite{Schreck_2011,Henkes_2012,Ikeda_2013,Goodrich2_2014,Mizuno2_2016,Otsuki_2014,Coulais_2014}.
Even the elastic regime shrinks and disappears on approach to the jamming transition $\varphi_c$.
Thus, to understand more generally the mechanical and vibrational properties of systems on the edge of marginal stability, inevitably it might be necessary to take into account non-linear effects, which should be distinct between different potentials, i.e., different values of $a$.

Finally, we highlight a prescient feature to our findings.
Our results show that the linear elastic response of the systems studied here reflects the nature of the vibrational spectrum.
Therefore, it is reasonable to expect that materials with different distributions of vibrational modes should exhibit different elastic responses~\cite{Schreck_2011,Koeze_2016,Cakir_2016}.
Given that the density of vibrational states is accessible through various scattering~\cite{Sette_1998,ruffle_2003,Monaco_2006,bruna_2011} and covariance matrix measurement~\cite{brito_2010,Ghosh_2010,Chen_2010,Henkes_2012} techniques, it should be possible to pin down the expected elastic behavior through such measurements.
Also, our results highlight the concept that materials of desired functionality or tunable mechanical behavior may be fashioned through adaptive manufacturing techniques whereby desirable constituent motifs are structured to achieve designer vibrational mode distributions.

\begin{acknowledgments}
This work was initiated at the Deutsches Zentrum f\"{u}r Luft- und Raumfahrt (DLR), Cologne for which L.E.S. greatly appreciates the support of the German Science Foundation DFG during a hospitable stay at the DLR under Grant No. FG1394.
We also thank Th.~Voigtmann, S.~Luding, M.~Otsuki, A.~Zaccone, and A.~Ikeda for useful discussions.
H.M. acknowledges support from DAAD (German Academic Exchange Service).
K.S. is supported by the NWO-STW VICI grant 10828.
\end{acknowledgments}

\bibliographystyle{apsrev4-1}
\bibliography{manuscript}

\end{document}